\documentclass[12pt]{article}
\usepackage{epsfig}
\usepackage{amsfonts}
\usepackage{amscd}
\usepackage{latexsym}
\usepackage{amsmath,amssymb}
\usepackage{verbatim}
\usepackage{color}
\usepackage{cite}
\usepackage[textheight=9in, textwidth=6.5in, letterpaper]{geometry}
\usepackage{soul}

\usepackage{hyperref}
\hypersetup{
    colorlinks,
    citecolor=black,
    filecolor=black,
    linkcolor=black,
    urlcolor=black
}

\usepackage{graphicx}
\usepackage{tikz}
\usepackage{tikz-3dplot}
\usetikzlibrary{calc}
\usetikzlibrary{positioning}
\usepackage{float}
\usepackage{subfig}
\usepackage[T1]{fontenc}
\usepackage{amsmath, amssymb, amsthm, fullpage}

\usepackage{doc}
\usepackage{ifthen, color, fancyvrb}
\usepackage{nextpage}

\usepackage{wrapfig}
\usepackage{parskip}
\usepackage{authblk}
\usepackage[makeroom]{cancel}
\usepackage{slashed}
\def\e{\epsilon }
\def\be{\begin{equation}}
\def\ee{\end{equation}}
\def\bz{{\bar z}}

\def\p{\partial}

\def\be{\begin{equation}}
\def\ee{\end{equation}}
\def\bea{\begin{eqnarray}}
\def\eea{\end{eqnarray}}
\def\<{\langle }
\def\>{\rangle}

\def\eps{\epsilon}

\def\D{{\Delta}}
\def\ve{{\varepsilon}}
\def\bh{\bar{h}}
\def\ca{\mathcal{A}}

\def\co{\mathcal{O}}
\def\cp{\mathcal{P}}
\begin{document}
\begin{titlepage}
\unitlength = 1mm
\ \\
\vskip 4cm
\begin{center}

{ \LARGE {\textsc{Celestial Operator Products of Gluons and Gravitons}}}

\vspace{0.8cm}
Monica Pate{$^{*\ddagger}$}, Ana-Maria Raclariu{$^*$}, Andrew Strominger{$^*$} and Ellis Ye Yuan{$^{*\dagger}$}
\vspace{1cm}

\begin{abstract} 
The operator product expansion (OPE)  on the celestial sphere of conformal primary gluons and gravitons is studied. Asymptotic symmetries 
imply recursion relations between products of operators whose conformal weights differ by half-integers. It is shown, for tree-level Einstein-Yang-Mills theory, that these recursion relations are so constraining that they completely  fix the leading celestial OPE coefficients in terms of the Euler beta function. The poles in the beta functions are associated with conformally soft currents. 
\end{abstract}

\end{center}
\vspace{4.5cm}
{$*$ \it  Center for the Fundamental Laws of Nature, Harvard University,
Cambridge, MA, USA}\\ 
{$\dagger$ \it Zhejiang Institute of Modern Physics, Zhejiang University,
Hangzhou, Zhejiang, China}\\
{$\ddagger$ \it Society of Fellows, Harvard University, Cambridge, MA, USA}
\end{titlepage}

\pagestyle{empty}
\pagestyle{plain}

\tableofcontents
\section{Introduction}
The subleading soft graviton theorem implies that any quantum theory of gravity in an asymptotically flat four-dimensional (4D) spacetime 
has an infinite-dimensional 2D conformal symmetry \cite{Cachazo:2014fwa,Kapec:2014opa}. This  symmetry acts on the celestial sphere at null infinity, with Lorentz transformations generating  the global $SL(2,\mathbb{C})$ subgroup \cite{Strominger:2017zoo}.   4D scattering amplitudes in a conformal basis transform like  a collection of  correlators in a 2D 
`celestial conformal field theory'. Properties of the so-defined celestial CFTs have been extensively studied and differ in ways which are not yet fully understood  from those of conventional CFTs. Celestial operator spectra were studied in \cite{Cheung:2016iub, Pasterski:2017kqt,Donnay:2018neh,Himwich:2019dug, Banerjee:2018gce, Banerjee:2019aoy, Fotopoulos:2019tpe, Law:2019glh}
and celestial scattering amplitudes in \cite{Pasterski:2016qvg, Lam:2017ofc, Banerjee:2017jeg, Pate:2019mfs, Adamo:2019ipt,Stieberger:2018edy,Schreiber:2017jsr,Pasterski:2017ylz,Nandan:2019jas, Fan:2019emx,  Stieberger:2018onx,Cardona:2017keg,Puhm:2019zbl,Guevara:2019ypd}.

In a general celestial CFT, the operator spectrum is continuous, with one continuum for every $stable$ species of particles. 
 Unstable particles decay before reaching infinity and are not part of the data on the celestial sphere. For a stable particle of spin $s$, a complete basis is given by celestial conformal primaries with  conformal weights $(h,\bar h)=({ \Delta+s\over2},{\Delta-s\over 2})$ and ${\rm Re}(\D)=1$  \cite{Pasterski:2017kqt}. 
 
 In this paper we study the operator product expansion (OPE) of these celestial primaries. 
 Poles in the celestial  OPE for massless particles turn out to be Mellin transforms of  collinear singularities in momentum space which can be computed with Feynman diagrams. The OPEs follow from the three-point vertices coupling  the stable particles.  We derive  a simple and universal  formula \eqref{smp} relating the conformal weights in the operator product expansion to the bulk scaling dimension of the three-point vertex. 

Celestial CFTs are subject to multiple infinities of asymptotic symmetry constraints beyond the familiar ones following from 
2D 	conformal symmetry. These constraints have no analogs in conventional CFTs. They follow from the leading and subsubleading soft graviton theorems and, if there are gauge bosons, the subleading soft 
photon/gluon theorem. On the face of it, it would seem impossible for a collection of celestial amplitudes to satisfy additional infinities of constraints, but of course we know this seemingly overconstrained problem must have a solution as many celestial amplitudes have been explicitly constructed. So far there has been little study of the implications of these constraints.

In this paper we show that the additional symmetry  constraints have remarkable implications for the operator product expansion.  They imply recursion relations between products of celestial operators whose conformal weights differ by a half-integer. We analyze in detail tree-level Einstein-Yang-Mills (EYM) theory and find  that the recursion relations, together with some analyticity assumptions,  are so powerful that they completely determine (at least)  all the conformal primary OPE coefficients of the leading poles in the operator product expansion. They are given by Euler beta functions (ratios  of Gamma functions) with arguments given by  the conformal weights. 
We check that  the direct but lengthier Feynman-diagrammatic computation yields the same beta functions. 

Inclusion of quantum, stringy or other corrections would introduce higher dimension terms into the effective action. These may alter both the three-point vertices and the (sub)subleading soft theorems, and hence the subleading terms in the OPEs in accord with the general formula \eqref{gfms} below. It will be interesting to study the symmetry constraints on OPEs in this more general context, as well as to extend the analysis beyond the leading poles. 

In a conventional (unitary, discrete) CFT, the operator  spectrum and the conformal primary OPEs fully determine the theory.
Should an analogous result hold in celestial CFT, it would suggest that complete quantum theories of gravity are determined by these symmetry-constrained OPE coefficients. These are far fewer in number than the number of possible terms in the effective Lagrangian. This resonates with similar findings in the amplitudes program \cite{Cachazo:2013hca, Cachazo:2013iea, Arkani-Hamed:2013jha, Dixon:2014xca, Rodina:2018pcb}.  It would be interesting to study further constraints among  these OPEs from  crossing symmetry. 

This paper is organized as follows. Section \ref{sec:pre} contains conventions and useful formulae. Section \ref{sec:3} begins with a general derivation  of the relation between the bulk dimension of the three-point couplings and the conformal weights of the OPE. 
Subsection \ref{sec:3.1} considers the gluon OPE poles in tree-level Yang-Mills (YM) theory. The subleading soft gluon 
theorem is shown to imply recursion relations among the OPE coefficients, with the overall normalization fixed by the leading soft gluon theorem. For the collinear pole terms, these are uniquely solved -- subject to certain falloffs at large operator dimension -- by Euler beta functions. Subsection \ref{g} derives similar results, invoking the subsubleading soft graviton theorem,  for the graviton OPEs in Einstein gravity, while \ref{gg} derives the EYM gluon-graviton OPEs. In section \ref{sec:4}, building on previous analyses of collinear limits of gravitons and gluons, we directly compute the collinear singularities in momentum space and then the OPE poles via a Mellin transformation. This direct analysis fully agrees with the symmetry-derived results. We generalize our results for operators associated to incoming and outgoing particles in section \ref{sec:in-out}.  The EYM OPEs are all summarized in section \ref{sec:3.4}. Appendix \ref{celOPE} details the relation between the bulk scaling dimension of a three-point vertex and the conformal weights entering the OPE. Appendix \ref{ft} presents the list of all OPE coefficients which can be generated by higher-dimension operators. In appendices \ref{ssga} and \ref{ssgta} we review the unbroken global symmetries which are related to the subleading soft gluon theorem and subsubleading soft graviton theorem and used to derive the recursion relations. In appendix \ref{beta} we solve the recursion relations for the beta function and spell out the regularity  conditions which make the solution unique.

\section{Preliminaries}

\label{sec:pre}

In this section we give our conventions for celestial scattering amplitudes and collect some useful formulae. 
	
	Celestial amplitudes $\widetilde \ca$ of massless particles are obtained from momentum-space amplitudes  $\ca$ (including the momentum-conserving delta function) by performing Mellin transformations with respect to the particle energies\cite{Pasterski:2016qvg,Pasterski:2017kqt}
	\be\label{csa}
		\widetilde \ca_{s_1\cdots s_n} (\Delta_1, z_1, \bz_1, \cdots, \Delta_n, z_n, \bz_n) 
			 = \left(\prod_{k = 1}^n \int_0^\infty d \omega_k ~ \omega_k^{\Delta_k -1}\right)  \ca_{s_1\cdots s_n} (\epsilon_1\omega_1, z_1, \bz_1,\cdots, \epsilon_n\omega_n, z_n,\bz_n) ,
	\ee
	where the helicity $s_k=\pm1$ for gluons and $s_k=\pm2$ for gravitons. In order to write momentum-space amplitudes as functions of $(\epsilon_k\omega_k, z_k, \bz_k)$, we parametrize the Cartesian coordinate massless 4-momenta components as
	\be \label{mompar}
		p_k^{\mu} 
		= \frac{\epsilon_k \omega_k}{\sqrt{2}}(1 + z_k \bz_k, z_k + \bz_k, -i(z_k - \bz_k), 1 - z_k\bz_k),
	\ee
	with $\mu=0,1,2,3,$  $\epsilon_k = \pm 1$ for outgoing and incoming momenta respectively and helicities are defined with respect to outgoing momenta. In the following two sections we compute OPEs of outgoing states with $\epsilon_k = 1$. We finally explain how to generalize the analysis to mixed incoming and outgoing OPEs in section \ref{sec:in-out}. Color indices and in/out labels on celestial amplitudes are suppressed.  We later use $A$ to denote color-ordered partial amplitudes. We note that 
	\be\label{pot} p_1\cdot p_2=-\epsilon_1 \epsilon_2 \omega_1 \omega_2 z_{12}\bz_{12},\ee
	where \be z_{12}=z_1-z_2, ~~~ \bz_{12}=\bz_1-\bz_2.\ee 
	
For coordinates
\be
\begin{split}
x^{\mu} &= u \p_z\p_{\bz} q^{\mu}(z, \bz) + r q^{\mu}(z,\bz), \\
q^{\mu}(z, \bz) &= \frac{1}{\sqrt{2}}\left(1 + z\bz, z + \bz, -i(z - \bz), 1- z\bz \right) ,
\end{split}
\ee
the flat metric is 
	\be ds^2 = dx^{\mu}dx_{\mu} = -2dudr+ 2r^2dzd\bz,\ee
the celestial sphere is conformally mapped to the celestial plane and $z_k$ is the spatial location at which a particle of momentum $p_k$ crosses ${\cal I}^+$.  $\widetilde \ca$ transforms as a correlator  of $n$  weight $(h_k,\bar h_k)=({ \Delta_k+s_k\over2},{\Delta_k-s_k\over 2})$ primaries under conformal transformations of the celestial plane.  In the next two sections we consider only OPEs between outgoing particles, and use 
	${\cal O}_{\D,s}$ to denote a generic such  primary, $O^{\pm a}_\D$ for a primary gluon where $s=\pm1$ (with $a$ an adjoint group index) and $G^{\pm }_\D$ for a primary graviton with $s=\pm2$. In section \ref{sec:in-out} we reintroduce the  additional label $\epsilon$ to distinguish between incoming and outgoing operators $\mathcal{O}_{\Delta, s}^{\epsilon}$. (Whenever the label is absent, the operator is taken to be outgoing.) Group structure constants $f^{ab}_{~~c}$ obey the Jacobi identity 
\be\label{dsm} f^{ab}_{\ \ d}f^{dce}+ f^{bc}_{\ \ d}f^{dae}+ f^{ca}_{\ \ d}f^{dbe} =0,
\ee
and generators are normalized such that
\be \label{gen_norm}
{\rm Tr}(T^a  T^b) = g_{YM}^2 \delta^{ab},
\ee	
where $T^a$ are in the fundamental representation.
We work with the following polarization vectors for massless spin-1 particles
	\be
	\label{pt}
		\ve_k^{+}{}^\mu =  { 1 \over \epsilon_k \omega_k}\p_{z_k}   p_k^{\mu}, \quad \quad \quad \ve_k^{-}{}^\mu =    { 1 \over \epsilon_k \omega_k}\p_{\bz_k} p_k^{\mu},
	\ee
	and polarization tensors $\ve_k^{\pm}{}^{ \mu\nu} =\ve_k^{\pm}{}^\mu \ve_k^{\pm}{}^\nu$ for massless spin-2 particles. These obey
	\be
	\label{ped} p_1\cdot\ve^{-}_2= \epsilon_1 \omega_1z_{12},~~~ p_1\cdot\ve^{+}_2=\epsilon_1\omega_1\bz_{12}.\ee
	
	Generically, the Mellin transform $\omega_k$-integrals converge only for restricted values of $\Delta_k$. For example in gauge theory they converge on the unitary principle series with ${\rm Re }(\D)=1$. However we will be interested in  the celestial amplitudes for other complex values of $\Delta_k$, where we define them by analytic continuation.

\section{OPEs from asymptotic symmetries}

\label{sec:3}

In this section we  study OPEs of conformal primary gluon and graviton operators on the celestial plane labeled by $(z, \bz)$. $z$ and $\bz$ will be varied independently. (These variables are independent  in (2,2) signature, for which the celestial plane becomes Lorentzian.)   Moreover we consider only the `holomorphic limit'  $z_{12}\to 0$ with $\bz_1, \bz_2$ fixed. Symmetry implies similar OPEs  for $\bz_{12}\to 0$  with $z_1, z_2$ fixed. However, order-of-limits subtleties arise when both $z_{12}\to0$ and $\bz_{12} \to0$ \cite{Stieberger:2015kia, He:2015zea, Distler:2018rwu}.  These are likely important for the structure of celestial amplitudes but are beyond the scope of this paper. 

Singularities in the celestial OPEs are the  Mellin transforms of collinear divergences in the momentum-space scattering amplitudes. 
This allows us to deduce some simple properties of the OPEs without any detailed computations.  Collinear singularities arise when  $p_1 || p_2$ for massless particles which  couple via a three-point vertex to form a nearly on-shell internal particle. The resulting propagator is proportional to  ${ 1 \over p_1\cdot p_2}$ which, according to \eqref{pot}, diverges as  ${1 \over z_{12}}$  for $z_{12} \to 0$. Hence two-operator  OPE singularities are at most simple poles in $z_{12}$. Schematically the OPE  of conformal primaries ${\cal O}_{\D,s}$ with conformal weights $(h,\bar h)=({ \Delta+ s\over2},{\Delta-s\over 2})$ takes the form 
\be\label{ope} {\cal O}_{\D_1,s_1}(z_1,\bz_1){\cal O}_{\D_2,s_2}(z_2,\bz_2)\propto { 1\over z_{12}}{\cal O}_{\D_3,s_3}(z_2,\bz_2)+{\rm order}\big(z_{12}^0\big).\ee 

Contributions to the OPE \eqref{ope} arise from the  three-point interaction vertices in the expansion of terms in the bulk effective Lagrangian around flat space. Since gravitons and gluons have bulk scaling dimension one, these are characterized by bulk
 dimension $d_V=3+m$, where $m$ is the number of spacetime derivatives. For example the most relevant gluon-gluon-graviton vertex $h\p A\p A$ has $d_V=5$, while the gluon-gluon-gluon vertex $A\p A A$ has $d_V=4$. 
The conformal weight $\Delta_3$ of the operator on the right hand side of \eqref{ope} can be inferred from  $d_V$.    Each  derivative leads to one extra factor of $\omega$ inside the Mellin transform \eqref{csa}, and therefore shifts  $\Delta_3$ up by one.  Accounting for all the factors of $\omega$ (including two in the internal propagator), one finds that the OPE of two operators of conformal weight $\D_1$ and $\D_2$ which couple via a three-point vertex of bulk dimension $d_V$ can only produce an  operator with conformal weight
\be \label{smp} \D_3=\D_1+\D_2+d_V-5.\ee  
Details are in appendix \ref{celOPE}. Further, insisting on conformal invariance, one finds that the contribution to the OPE from a vertex of fixed $d_V$ must take the form\footnote{ Since there are no gauge and coordinate invariant $d_V<4$ relevant operators in a theory with only gluons or gravitons (except of course the cosmological constant, which we assume vanishes!), there are no $ \frac{1}{z_{12}\bz_{12}}$ singularities.\vspace{3pt}}\be 
\begin{split}
\label{gfms} {\cal O}_{\D_1,s_1}&(z_1,\bz_1){\cal O}_{\D_2,s_2}(z_2,\bz_2)\\
&\sim
\sum_{n = 0}^{d_V -  4} c_{n,  d_V} (\D_1,s_1;\D_2,s_2) z_{12}^{n-1} \bz_{12}^{d_V - 4 - n}\mathcal{O}_{\Delta_1 + \Delta_2 + d_V - 5, s_1 + s_2 + 3 + 2n - d_V}(z_2,\bz_2).
\end{split}
\ee
Although in the most general case \eqref{gfms} is an infinite series when summing over $d_V$, many terms are eliminated  when the spins range over limited values. For example in a theory with only $s=\pm1$ 
gluons the $O^+O^+$ OPE in \eqref{gfms} reduces to the two terms\footnote{Due to the lower limit in \eqref{gfms}, the second term is absent for $d_V = 4$.} 
\be
\label{gope}
\begin{split}
O^{+a}_{\Delta_1}(z_1,\bz_1) O^{+b}_{\Delta_2}(z_2,\bz_2) &\sim i f^{ab}_{~~c}\left(c_{\frac{d_V}{2} - 2, d_V}z_{12}^{d_V/2 - 3}\bz_{12}^{d_V/2 - 2}O^{+c}_{\Delta_1 + \Delta_2 + d_V - 5}(z_2,\bz_2)\right. \\
&\qquad\qquad\quad  \left. +\  c_{\frac{d_V}{2} - 3, d_V} z_{12}^{d_V/2 - 4} \bz_{12}^{d_V /2-1}O^{-c}_{\Delta_1 + \Delta_2 + d_V - 5}(z_2,\bz_2)\right).
\end{split}
\ee
In this paper we consider in detail only symmetry constraints on the leading ($n=0$) pole terms in EYM theory, for which there are only seven nonzero coefficients $c_{0,d_V}$ with $d_V = 4,5$. These are all completely fixed by asymptotic symmetries and  summarized in section \ref{sec:3.4}. 
Equally powerful symmetry constraints apply to all terms in the expansion (\ref{gfms}), but the  more intricate higher-order analysis is left to future investigation. 

\subsection{Gluons}
\label{sec:3.1}

In this section we consider  pure renormalizable glue theory with $d_V=4$. In this case,\footnote{ Additional terms on the right hand side in the presence of gravitons are determined in subsection \ref{gg}.  An $F^3$ term with $d_V=6$ would lead to an $O^-$ term on the right hand side of \eqref{saz}. }
 \be \label{saz}
O^{+a}_{\D_1}(z_1,\bz_1) O^{+b}_{\D_2}(z_2, \bz_2) \sim -\frac{i f^{ab}_{~~c}}{z_{12}} C (\D_1, \D_2) O^{+c}_{\D_1 + \D_2-1}(z_2, \bz_2),
 \ee
 \be \label{sbz}
O^{+a}_{\D_1}(z_1,\bz_1) O^{-b}_{\D_2}(z_2, \bz_2) \sim -\frac{i f^{ab}_{~~c}}{z_{12}} D (\D_1, \D_2) O^{-c}_{\D_1 + \D_2-1}(z_2, \bz_2),
 \ee
 for some to-be-determined coefficients  $C (\D_1, \D_2) =C (\D_2, \D_1) $ and $D (\D_1, \D_2)$. $O^{-}O^{-}$ is nonsingular in $z_{12}$. For gluons, the conformal primaries  with ${\rm Re }(\D)=1$ are a complete basis of square-integrable wave packets \cite{Pasterski:2017kqt}. We see that in the renormalizable theory the  OPEs \eqref{saz}, \eqref{sbz} close on such operators.   

The OPE coefficients are  subject to a number of symmetry constraints. The simplest is  translations $\mathcal{P} $ towards the `north pole' of the celestial sphere, which involves a factor of $\omega$ in momentum space. In a conformal basis, this symmetry shifts the operator dimension \cite{Donnay:2018neh,Stieberger:2018onx}:
\be
\label{Psym}
 \delta_{\mathcal{P}} O^{\pm a}_{\D}( z,\bz) = O^{\pm a}_{\Delta + 1}(z, \bz).
 \ee
Acting on both sides of \eqref{saz} and \eqref{sbz}  with $\delta_{\mathcal{P}}$ gives the recursion relations
\be \label{rec}  C (\D_1, \D_2)=C (\D_1+1, \D_2)+C (\D_1,\D_2+1),\ee
\be \label{rdc}  D (\D_1, \D_2)=D(\D_1+1, \D_2)+D (\D_1,\D_2+1).\ee
Such relations were also found in \cite{Law:2019glh}.

Next, the leading conformally soft theorem is \cite{Fan:2019emx,Pate:2019mfs,Nandan:2019jas}
\be \label{sabz}
\lim_{\D_1\to 1}O^{+a}_{\D_1}(z_1, \bz_1) O^{\pm b}_{\D_2}(z_2, \bz_2) \sim -\frac{i f^{ab}_{~~c}}{(\D_1-1)z_{12}}  O^{\pm c}_{\D_2}(z_2, \bz_2).
 \ee
This implies poles in $C $ and $D$ with residues
\be \label{bc1}\lim_{\D_1\to 1}(\Delta_1 - 1)C (\D_1, \D_2)= \lim_{\D_1\to 1}(\Delta_1 - 1)D (\D_1, \D_2) =  1.\ee

Further, less familiar,   constraints come from the subleading soft symmetry  parametrized by   ($Y^{za}, Y^{\bz a}$).
Under these symmetries, the gauge field on $\cal{I}^+$ shifts  by
\cite{Lysov:2014csa}
\be \label{kh} \delta_Y A_z^a=  u\p_z^2Y^{za},~~~~~ \bar{\delta}_{Y} A_\bz^a=  u\p_\bz^2Y^{\bz a}.\ee
If the right hand side is nonzero, the symmetry is spontaneously broken. The unbroken symmetries are the most useful for present purposes. These correspond to 
$Y^{z a}=z\e^a, \e^a$ and  $Y^{\bz a}=\bz\e^a, \e^a$
for constant $\e^a$.  As shown in appendix \ref{ssga} (see also \cite{Lysov:2014csa}), for the global symmetry $Y^{z a}=z\e^a$ 
conformal primary gluons transform as 
\be \label{subsoftglue+}\delta_b O^{\pm a}_{\D}(z,\bz )=-(\D-1\pm1 +z\p_z) if^{a}_{~bc}O^{\pm c}_{\D-1}(z,\bz ).\ee
Similarly for $Y^{\bz a}=\bz \e^a$ we have 
\be\label{rdf} \bar \delta_b O^{\pm a}_{\D}(z,\bz )= -(\D-1\mp1 +\bz\p_\bz) if^{a}_{~bc}O^{\pm c}_{\D-1}(z,\bz ).\ee
Since they are unbroken, the Ward identities for these symmetries involve no soft insertions
\be \label{sh}\begin{split}\sum_{ k=1}^n\langle O_1\cdots \delta O_k\cdots O_n\rangle=0,\qquad \qquad \sum_{k =1}^n\langle O_1\cdots \bar \delta O_k\cdots O_n\rangle=0.\end{split}\ee

We now extract the  consequences of this global symmetry  for the  OPE \eqref{saz}. This is complicated by the appearance of derivatives in the transformation laws \eqref{subsoftglue+} and \eqref{rdf} which mix up primaries and descendants, and therefore do not map the leading OPE relations \eqref{saz} and \eqref{sbz} to themselves. These bothersome terms can be eliminated in $\bar \delta$ by considering the special case $\bz_1=\bz_2=0$, 
where \eqref{saz} still holds.  (The $z$-analog of this trick cannot be used to analyze the implications of $\delta$ symmetry because \eqref{saz} blows up for $z_1=z_2=0$.) 
Acting with $\bar \delta_d $ on both sides of \eqref{saz} we get 
\be \label{szaz}
\begin{split}
(\D_1-2)if^a_{~dc}O^{+c}_{\D_1-1}(z_1,0) O^{+b}_{\D_2}( z_2,0) + (\D_2-2)O^{+ a}_{\D_1}(z_1,0)i f^b_{~dc}O^{+ c}_{\D_2-1}( z_2,0) \\  \sim \frac{\D_1+\D_2-3}{z_{12}} C (\D_1, \D_2) f^{ab}_{~~c}f^c_{~de}O^{+e}_{\D_1 + \D_2-2}(z_2,0).
\end{split}
 \ee
Using the OPE again on the left hand side we obtain the consistency condition
\be \label{szz}
\begin{split}
(\D_1-2)C(\D_1-1,\D_2) f^a_{~dc}f^{cb}_{~~e} + (\D_2-2) C(\D_1,\D_2-1)f^b_{~dc}f^{ac}_{~~e} \\ = (\D_1+\D_2-3)C (\D_1, \D_2) f^{ab}_{~~c}f^c_{~de}.
\end{split}
 \ee
Applying the Jacobi identity \eqref{dsm}
this implies
\be \label{sbdz} (\D_1-2)C(\D_1-1,\D_2)= (\D_1+\D_2-3)C (\D_1, \D_2). \ee
Under suitable assumptions spelled out in appendix \ref{beta} about boundedness and  analyticity in $\Delta_1, \Delta_2$ (basically that there are no poles other than those implied by the soft theorems), \eqref{sbdz} together with the normalization condition \eqref{bc1} have the unique solution\footnote{ Symmetry of $C(\Delta_1, \Delta_2)$ under $\Delta_1 \leftrightarrow \Delta_2$ together with the subleading soft symmetry constraint \eqref{sbdz} in fact imply the translation invariance relation \eqref{rec}, which therefore does not further constrain $C$.} 
\be  \label{solO+O+}C (\D_1,\D_2)=B(\D_1-1,\D_2-1)  ,\ee
where $B$ is the Euler beta function
\be B(x,y)= {\Gamma(x)\Gamma(y)\over \Gamma(x+y)}.\ee
Acting with $\bar \delta_d$ on both sides of \eqref{sbz} gives a slightly different result because of the $\pm$ in \eqref{rdf}. Instead of \eqref{sbdz}
we find two different recursion relations
\be \label{scdz} 
\begin{split} 
(\Delta_1 - 2) D(\Delta_1 - 1, \Delta_2) = (\D_1 + \D_2 - 1) D(\D_1, \D_2),\\
\D_2D(\D_1,\D_2-1)= (\D_1+\D_2-1)D(\D_1, \D_2).
\end{split} \ee
Again, \eqref{scdz} together with the normalization condition \eqref{bc1}, have the unique solution \be  \label{solO+O-} D (\D_1,\D_2)=B(\D_1-1,\D_2+1)  .\ee
\eqref{solO+O+} and \eqref{solO+O-} agree with the  expressions  previously obtained in \cite{Fan:2019emx} by direct Mellin transform of the collinear singularities in momentum space. Here we see the OPE is entirely fixed by symmetries. 

In fact there are further consistency conditions, which we did not need to use to fix $C$ and $D$, but it can be checked that they are satisfied.  One of these is that the OPEs have properly normalized poles at $\D_1\to 0$ corresponding to the subleading soft theorem. This is indeed manifest in 
\eqref{solO+O+} and \eqref{solO+O-}.  We have used here only a few global symmetries. There are infinitely many more  constraints from the infinity of soft symmetries. However these may all be obtained by commuting the global symmetries with the local conformal symmetry, which is manifestly built in to our construction and so their satisfaction is guaranteed.  

\subsection{Gravitons}
\label{g}

For gravitons in Einstein gravity the three-point vertex has $d_V=5$. According to \eqref{gfms} this leads to an OPE of the form\footnote{A contribution of the form
 $ \frac{\bz_{12}^5}{z_{12}}  E'_+ (\Delta_1, \Delta_2) G^{-}_{\Delta_1 + \Delta_2+4}$  to the $G^+_{\Delta_1}G^+_{\Delta_2}$ OPE might for example be 
 generated  by an $R^3$ correction to the Einstein-Hilbert action.  } 
 \be \label{gsaz}
\begin{split}
G^+_{\D_1}(z_1,\bz_1) G^{\pm}_{\D_2}(z_2,\bz_2) &\sim \frac{\bz_{12}}{z_{12}}E_{\pm} (\D_1, \D_2) G^\pm_{\D_1 + \D_2}( z_2,\bz_2),
\end{split}
 \ee
for some to-be-determined coefficients  $E_+ (\D_1, \D_2) = E_+(\D_2, \D_1)$ and $E_-(\D_1, \D_2)$, while $G^-G^-$ is nonsingular in the $z_{12} \to 0$ limit.
As for the case of  gluons, translation invariance  implies the recursion relation
\be \label{grec}  E_{\pm} (\D_1, \D_2)=E_{\pm} (\D_1+1, \D_2)+E_{\pm} (\D_1,\D_2+1).\ee
The residue of a pole at $\D_1\to 1$ is fixed by the 
the leading  soft graviton theorem\footnote{Supertranslations are generated by the current 
$ P_z=-\frac{2}{\kappa}\lim_{\Delta\to 1}(\D-1)\p_\bz G_\D^+$.}\cite{Puhm:2019zbl}
\be  \label{leadgrav}\lim_{\D_1\to 1}E_{\pm} (\D_1, \D_2)\sim -{\kappa \over 2(\D_1-1)}, \qquad \kappa = \sqrt{32\pi G}.\ee

The subleading soft symmetry corresponds to 2D conformal transformations, which are generated by the shadow of $G^+_0$ \cite{Kapec:2016jld, Kapec:2017gsg, Donnay:2018neh, Fotopoulos:2019tpe}. However, by working in a conformal basis, we have already  ensured that 
the OPE is conformally invariant, and no further constraints on $E_\pm$ are obtained from the subleading soft symmetry.  

The role of the subleading soft gluon theorem in constraining gauge theory OPEs is here played by the $subsubleading$ soft graviton theorem,
which implies further global symmetries. We show in appendix \ref{ssgta} that the relevant gravitational analog of the gauge theory relation \eqref{rdf} is
\be
\label{sssc}
\begin{split}
 \bar \delta G^{\pm}_{\D}(z,\bz) &= -\frac{\kappa}{4} \left[(\D\mp 2)(\D \mp 2 - 1) +4(\D \mp 2)\bz \p_{\bz} + 3 \bz^2\p_{\bz}^2\right]G^{\pm}_{\D-1}(z,\bz). 
 \end{split}\ee
 However, to study the consequences of this symmetry on the OPE, we cannot directly set $\bz_1=\bz_2=0$ in \eqref{gsaz} because that will set the right hand side to zero and no useful relation would be obtained. To avoid this we first differentiate with respect to $\bz_1$, and then set $\bz_1=\bz_2=0$. The positive helicity graviton OPE in \eqref{gsaz} is then
 \be \label{gsaz1b}
\p_{\bz_1}G^+_{\D_1}(z_1,0) G^+_{\D_2}(z_2,0) \sim \frac{E_+(\D_1, \D_2)}{z_{12}} G^+_{\D_1 + \D_2}(z_2,0).
 \ee
Equation \eqref{sssc} becomes
 \be
\label{sssc1}
 \bar \delta G^+_{\D}(z,0) = - \frac{\kappa}{4}(\D-2)(\D-3)G^+_{\D-1}( z,0) ,\ee
 and in addition implies 
 \be
 	\bar \delta \p_{\bz}  G^+_{\D}(z,0) =- \frac{\kappa}{4} (\D-2)(\D+1)\p_{\bz}G^+_{\D-1}(z,0).
 \ee  Invariance of the OPE \eqref{gsaz} then holds if and only if
\be\label{gsbdz}
\begin{split}
(\D_1 +1)(\D_1- 2)E_{\pm} (\D_1-1,\D_2)&+(\D_2 \mp 2 -1)(\D_2 \mp 2)E_{\pm} (\D_1,\D_2-1)\\
&=(\D_1+\D_2\mp 2)(\D_1+\D_2\mp 2 - 1)E_{\pm} (\D_1, \D_2).
\end{split} \ee
The two  recursion relations 
 \eqref{grec} and \eqref{gsbdz}, together with the normalization condition \eqref{leadgrav} are again solved by  Euler beta functions
 \be 
 \begin{split} \label{grpp} E_{\pm} (\D_1,\D_2)= -\frac{\kappa}{2} B(\D_1-1,\D_2\mp 2 + 1)
 \end{split}  .\ee
In section \ref{ggOPE} (see equations \eqref{gravOPEpp}, \eqref{gravOPEpm}) we directly compute  the Mellin transform of the  near-collinear graviton amplitudes and find complete agreement with 
 \eqref{grpp}. 

Additionally, the OPE coefficients $E_\pm$ must have properly normalized poles at $\Delta_1 \to 0$ and $\Delta_1 \to -1$ associated to the subleading and subsubleading soft graviton symmetries, respectively.  As
 in the gauge theory case, we did not impose such conditions in our derivation, but find that our results are consistent with these conditions.

\subsection{Gravitons and Gluons}
\label{gg}

In this section we consider OPEs involving both gravitons and gluons. The Einstein-Yang-Mills interaction (schematically $hF^2$) has $d_V=5$. The relevant term in \eqref{gfms} is \be \label{gsaz1}
G^+_{\D_1}(z_1, \bz_1 ) O^{\pm a}_{\D_2}(z_2, \bz_2 ) \sim \frac{\bz_{12}}{z_{12}} F_\pm (\D_1, \D_2) O^{\pm a}_{\D_1 + \D_2}(z_2, \bz_2).
 \ee  Translation invariance again  implies the recursion relation \eqref{rec} for $F_\pm$.  A second set of relations is determined from 
 the global symmetry associated to subsubleading soft graviton theorem, whose action on gluons is shown in appendix \ref{ssgta} to be
  \be
  \label{sssggl}
 	\begin{split}
		\bar \delta O^{\pm a}_\Delta (z, \bz)& = -\frac{\kappa}{4} \left[(\D\mp 1 -1 )(\D \mp1 ) +4(\D\mp1 )\bz \p_{\bz} + 3 \bz^2\p_{\bz}^2\right] O^{\pm a}_{\Delta-1} (z, \bz).
	\end{split}
 \ee
Consistency of the OPE with this symmetry requires   
  \be\label{gsbdz1}
  \begin{split}
(\Delta_1 + 1)(\Delta_1 - 2)  F_\pm(\Delta_1 -1, \Delta_2) &+ (\Delta_2\mp 1 -1) (\Delta_2 \mp 1)F_\pm(\Delta_1, \Delta_2 - 1)\\
& = (\Delta_1 + \Delta_2 \mp 1-1)(\Delta_1 + \Delta_2\mp1) F_\pm(\Delta_1, \Delta_2),
 \end{split}
  \ee  
 where these relations are derived by studying the OPE of $\p_{\bz_1 }G^+_{\D_1}(z_1, 0) O^{\pm a}_{\D_2}(z_2, 0)$ as  in the previous section.
  Fixing the normalization with the leading soft graviton theorem one  finds\be \label{rds}
 \begin{split} F_{\pm} (\D_1,\D_2)&= - \frac{\kappa}{2} B(\D_1-1,\D_2 \mp 1+ 1).
  \end{split}\ee
  
In the presence of gravitons, the right hand side of the gluon OPE \eqref{sbz} can also receive a correction of the form
 \be\label{dsx}
 	\begin{split}
 	O^{+a}_{\D_1}&(z_1,\bz_1) O^{-b}_{\D_2}(z_2, \bz_2) \\&\sim  -\frac{i f^{ab}_{~~c}}{z_{12}} B (\D_1-1, \D_2+1) O^{-c}_{\D_1 + \D_2-1}(z_2, \bz_2)+  \delta^{ab}  \frac{\bz_{12}}{z_{12}}  H(\Delta_1, \Delta_2) G^-_{\D_1 + \D_2}(z_2, \bz_2),
	\end{split}
 \ee
 corresponding to the fact that two gluons can make a graviton. This new term might seem to violate the subleading soft gluon theorem. Indeed, we will  find  shortly that symmetry constrains $H$ to have a pole associated with the subleading soft gluon symmetry at $\Delta_1 = 0$. However, as shown in \cite{Elvang:2016qvq,Laddha:2017vfh}, this theorem is corrected at tree-level in Einstein-Yang-Mills theory by the $hF^2$ coupling! The known form of the correction in fact can be used to fix the constant normalization of $H$. 
  
 Translation invariance implies $H$ obeys a recursion relation of the form  \eqref{rec}, while the subsubleading soft graviton theorem implies $H$  obeys the recursion relation
 \be
 	\begin{split}
		(\Delta_1 +2) (\Delta_1 -1) H (\Delta_1-1, \Delta_2)+ \Delta_2& (\Delta_2 +1)H(\Delta_1, \Delta_2-1) \\
			&= (\Delta_1 +\Delta_2 +2)(\Delta_1 +\Delta_2 +1) H(\Delta_1, \Delta_2).
	\end{split}
 \ee
The properly normalized solution is 
 \be
 \label{H}
 	H (\Delta_1, \Delta_2) = \frac{\kappa}{2} B (\Delta_1, \Delta_2 +2 ).
 \ee
The symmetry-derived results  \eqref{rds} and \eqref{H}
 agree with the Mellin transforms of direct Feynman diagram computations found in the next section. 
 
 The appearance of a graviton in the OPE of two gluons is presumably the boundary manifestation of the still-enigmatic 
 double-copy relation \cite{Kawai:1985xq,Bern:2008qj,Bern:2010ue}, in which gravity is the square of gauge theory. A remarkable discovery due to Stieberger and Taylor \cite{Stieberger:2014cea, Stieberger:2015kia}
is  that a pair of collinear gluons in a scattering amplitude can be replaced by a single graviton. If we take $\D_1=\D_2=0$ in \eqref{dsx}, the right hand side contains $G^-_0$ which is the shadow of the boundary stress tensor. This is a Sugawara-like construction of the stress tensor from a pair of subleading  soft currents. We leave these fascinating connections to future exploration. 
 
\section{OPEs from collinear singularities} \label{sec:4}
In this section we directly compute the celestial OPEs among gravitons and gluons in EYM by Mellin transforms of  Feynman diagrams.  We begin by reviewing the collinear limits of gauge and gravity amplitudes. The various OPEs are derived by Mellin transforming the corresponding amplitudes in the collinear limit and found in  all cases to agree with the symmetry-inferred results summarized later in section \ref{sec:3.4}. The OPEs among gluons were already derived in this manner in \cite{Fan:2019emx}. Their computation confirms  \eqref{solO+O+} and \eqref{solO+O-} and will not be repeated here. 

\subsection{Gravitons}
\label{ggOPE}

The collinear limits of gravity amplitudes were first derived in \cite{Bern:1998sv} and further developments are in \cite{White:2011yy,Akhoury:2011kq}. The leading divergence is generically protected against loop corrections\cite{Bern:1998sv}.   Here we specialize to a holomorphic collinear limit.

Consider a tree-level $n$-graviton scattering amplitude. In the limit when $z_{ij} \rightarrow 0$ for fixed $\bz_i, \bz_j$, the amplitude contains a universal
piece which factorizes as
\be
\begin{split}
\lim_{z_{ij} \rightarrow 0}\mathcal{A}_{s_1 \cdots  s_n}(p_1,\cdots, p_n) \longrightarrow \sum_{s = \pm 2}{\rm Split}_{s_i s_j}^{s}(p_i, p_j) \mathcal{A}_{s_1 \cdots s \cdots s_n}(p_1,\cdots ,P,\cdots,p_n),
\end{split}
\ee
where in the collinear limit\footnote{At subleading order in $z_{ij}$, \eqref{P} receives corrections, but these do not affect the leading singularities considered here.  For a discussion of subleading terms see \cite{Stieberger:2015kia}.}
\be
\label{P}
P^\mu = p_i^\mu + p_j^\mu, 
\quad \omega_P = \omega_i + \omega_j.
\ee
The collinear factor ${\rm Split}^{s}_{s_i s_j}(p_i, p_j)$  then takes the form\cite{Bern:1998sv}\footnote{We work with the Einstein-Hilbert action normalized as $S = \frac{2}{\kappa^2}\int d^4 x\sqrt{-g}R, \ \ g_{\mu\nu} = \eta_{\mu\nu} + \kappa h_{\mu\nu}$. This yields the following leading soft factor $S^{\pm}_{(0)} = \frac{\kappa}{2}\sum_k \frac{(p_k\cdot \varepsilon^{\pm})^2}{p_k\cdot q}$.}
\be
	\begin{split}
	&{\rm Split}^{2}_{22}(p_i, p_j) = -\frac{\kappa}{2} \frac{\bz_{ij}}{z_{ij}} \frac{\omega_P^2}{\omega_i \omega_j}, \qquad 
				{\rm Split}^{-2}_{2-2}(p_i, p_j) =	 -\frac{\kappa}{2} \frac{\bz_{ij}}{z_{ij}} \frac{\omega_j^3}{\omega_i \omega_P^2} ,\qquad
	\end{split}
\ee
with all other combinations of helicities vanishing. 
 In the collinear limit, the celestial gravity amplitude $\widetilde \ca$  becomes
\be
\begin{split}
\label{colll}
\widetilde{\mathcal{A}}_{s_1\cdots s_n} &(\Delta_1,z_1,\bz_1, \cdots, \Delta_n,z_n,\bz_n) \stackrel{i || j}{\longrightarrow}\\
 &\prod_{k = 1}^n   \int_0^{\infty} d\omega_k~ \omega_k^{\Delta_k-1}   \sum_{s = \pm 2}{\rm Split}^{s}_{s_i s_j}(p_i, p_j) \mathcal{A}_{s_1\cdots s\cdots s_n}(p_1,\cdots, P,\cdots,p_n)+ \cdots .
 \end{split}
\ee
To simplify, we make the following change of variables, 
\be	
		\omega_i = t \omega_P, \quad \quad \quad \omega_j = (1-t) \omega_P,
	\ee
	so that for example
	\be
		\int_0^{\infty} d \omega_i  \omega_i^{\Delta_i-1} \int_0^{\infty} d \omega_j  \omega_j^{\Delta_j-1} {\rm Split}^{2}_{22}(p_i, p_j) 
			 = -\frac{\kappa}{2} \frac{\bz_{ij}}{z_{ij}}\int_0^1 dt~  t^{\Delta_i-2} (1-t)^{\Delta_j-2} \int_0^\infty d \omega_P ~ \omega_P^{\Delta_i + \Delta_j -1}.
	\ee
The $t$ integral is immediately recognizable as the integral representation of the Euler beta function, 
\be \label{betafunction} B(x,y)= \int_0^1 dt~  t^{x-1} (1-t)^{y-1} ,\ee whose origin is hence  a splitting factor  for the conformal weight between the two collinear external  particles. Since the only $t$ dependence on the right hand side of \eqref{colll} comes from ${\rm Split}^{s}_{s_is_j}(p_i, p_j)$, one finds 
 \be \label{betaintegral042}
\begin{split}
	\lim_{z_{ij}\rightarrow 0} & \widetilde{\mathcal{A}}_{s_1\cdots 2\cdots 2\cdots} (\Delta_1,z_1,\bz_1, \cdots, \Delta_i, z_i, \bz_i, \cdots, \Delta_j, z_j,\bz_j, \cdots)  \longrightarrow  \\
	&  - \frac{\kappa}{2}\frac{\bz_{ij}}{z_{ij}} B(\Delta_i -1, \Delta_j -1) \widetilde{\mathcal{A}}_{s_1\cdots 2\cdots }(\Delta_1,z_1, \bz_1, \cdots, \Delta_i + \Delta_j,z_j, \bz_j ,\cdots) + {\rm order}(z_{ij}^0).
	\end{split}
\ee
Since this holds in any celestial amplitude, it implies the leading OPE between two positive helicity gravitons is
\be 
\label{gravOPEpp}
G^{+}_{\Delta_1}(z_1, \bz_1)G^{+}_{\Delta_2}(z_2, \bz_2) \sim  - \frac{\kappa}{2} \frac{\bz_{12}}{z_{12}} B(\Delta_1 - 1, \Delta_2-1) G^{+}_{\Delta_1 + \Delta_2}(z_2,\bz_2),
\ee
in agreement with \eqref{grpp}. By similar arguments, one also finds the following leading OPE between opposite helicity gravitons
\be \label{gravOPEpm}
		\begin{split} G^+_{\Delta_1}(z_1, \bz_1)G^-_{\Delta_2} (z_2 , \bz_2) & \sim   - \frac{\kappa}{2} \frac{\bz_{12}}{z_{12}}  B(\Delta_1-1, \Delta_2+3) G^-_{\Delta_1+\Delta_2} (z_2, \bz_2) ,
					\end{split}
	\ee
again in agreement with \eqref{grpp}. 
\subsection{Gravitons and gluons}
\label{sec:gr-gl}

In order to derive graviton-gluon OPEs from collinear limits of EYM amplitudes, we here derive the collinear limits of conventional momentum-space amplitudes.

We start with the general Stieberger-Taylor formula which relates a momentum-space amplitude of $n$ gluons and one graviton to a sum over color-ordered partial amplitudes of $n+1$ gluons\cite{Stieberger:2016lng}\footnote{Note that $\varepsilon^{\pm}(p)\cdot \chi_n = -\varepsilon^{\pm}(p)\cdot p =  0$ by momentum conservation, hence the sum in \eqref{stieberg-taylor} can be taken from $1$  to $n$. }
			\be \label{stieberg-taylor}
				A_{s_1\cdots s_n; \pm 2}(p_1, \cdots, p_n; p) = -\frac{\kappa}{2}\sum_{\ell = 1}^{n-1} (\varepsilon^\pm(p) \cdot \chi_\ell) A_{s_1\cdots s_\ell  ~\pm 1 ~s_{\ell+1}\cdots s_n}(p_1, \cdots, p_{\ell}, p, p_{\ell+1}, \cdots, p_n),
			\ee
		where $p_i,\  i = 1,...,n$ are the momenta of the gluons, $p$ is the momentum of the graviton, $\varepsilon(p)$ is the polarization of a gluon of momentum $p$  and
			\be \label{defchi}
				\chi_\ell = \sum_{k = 1}^\ell p_k.
			\ee
This formula allows us to determine  collinear graviton-gluon limits from collinear gluon limits. The known leading collinear behavior of gluon amplitudes arises from adjacent gluons in color-ordered partial amplitudes\cite{altarelli1977asymptotic}
\be
\begin{split}
\lim_{z_{ij}\rightarrow 0}A_{s_1\cdots s_n}(p_1,\cdots, p_i, p_j,\cdots,p_n) &\longrightarrow \sum_{s = \pm 1}{\rm Split}^{s}_{s_i s_j}(p_i, p_j) A_{s_1\cdots s \cdots s_n}(p_1,\cdots, P,\cdots,p_n),
\end{split}
\ee
where $P$ was defined in \eqref{P} and the non-vanishing ${\rm Split}^{s}_{s_i s_j}(p_i, p_j)$ for collinear gluons are given by
\be
\label{gl-split}
	\begin{split}
	&{\rm Split}^{1}_{11}(p_i, p_j) = \frac{1}{z_{ij}} \frac{\omega_P}{\omega_i \omega_j}, \qquad
				{\rm Split}^{-1}_{1-1}(p_i, p_j) =	 \frac{1}{z_{ij}} \frac{\omega_j}{\omega_i \omega_P} .
	\end{split}
\ee
Consider the collinear limit between a positive helicity  gluon of momentum $p_i$ and a positive helicity graviton.  In the collinear limit, the leading order contributions  from the right hand side of \eqref{stieberg-taylor}
are just the two terms where the gluon of momentum $p$ which replaces the graviton is adjacent to the $i^{\rm th}$ gluon:
\be
	\begin{split}
		\lim_{z_{i} - z\rightarrow 0} &A_{s_1\cdots 1\cdots s_n;2}(p_1, \cdots,p_i, \cdots,  p_n; p) \\ \longrightarrow &
		 -\frac{\kappa}{2}\left[ \left(   \varepsilon^+(p)\cdot   \chi_{i-1}\right) A_{s_1\cdots s_{i-1} ~1~ s_i\cdots s_n}(p_1, \cdots, p_{i-1}, p, p_{i}, \cdots, p_n)\right. \\
			 &\qquad \qquad \left. +	\left( \varepsilon^+(p) \cdot       \chi_i \right) A_{s_1\cdots  s_{i}~1~s_{i+1}\cdots s_n}(p_1, \cdots, p_i,p,p_{ i+1}, \cdots, p_n) \right]
				\\ \longrightarrow & \  -\frac{\kappa}{2}\varepsilon^+(p) \cdot 
		\left( \chi_i  -  \chi_{i-1}        \right) ~\frac{1}{z_i - z} \frac{\omega_P}{\omega_i \omega} A_{s_1\cdots s_{i-1}~ 1~s_{i+1}\cdots s_n}(p_1, \cdots,p_{i-1}, P,p_{i+1}, \cdots, p_n).
	\end{split}
\ee
We use \eqref{defchi} to further simplify
\be
	 \chi_i-  \chi_{i-1}  = p_i
\ee
and using \eqref{ped}, 
\be
	 \varepsilon^+(p) \cdot 
		\left( -  \chi_{i-1} +      \chi_i  \right)= \varepsilon^+(p) \cdot p_i = \omega_i (\bz_i- \bz).
\ee

Putting it all together, we obtain the following collinear limit for a positive helicity gluon and graviton 
\be
\begin{split}
\lim_{z_{i}-z \rightarrow 0}  A_{s_1\cdots 1\cdots s_n;2}&(p_1, \cdots,p_i,\cdots, p_n; p) \longrightarrow \\
					& -\frac{\kappa}{2}\frac{\bz_i - \bz}{z_i -z} \frac{\omega_P}{\omega}A_{s_1\cdots s_{i-1}~1~ s_{i+1}\cdots s_n}(p_1, \cdots p_{i-1} ,P, p_{i+1}\cdots, p_n).
					\end{split}
\ee

			By similar arguments, keeping only singular terms in $z_{i}-z$, we obtain the following collinear graviton-gluon limit for the mixed helicity case
			\be
				\begin{split}
				\lim_{z_{i} - z \rightarrow 0}	A_{s_1\cdots -1\cdots s_n;2}&(p_1, \cdots,p_i,\cdots, p_n; p) \longrightarrow \\
						&- \frac{\kappa }{2} \frac{\bz_i - \bz}{z_i -z} \frac{\omega_i^2}{\omega \omega_P}A_{s_1\cdots s_{i-1}~-1~s_{i+1}\cdots s_n}(p_1, \cdots p_{i-1},P,p_{i+1}\cdots, p_n).
									\end{split}
			\ee
		 Taking Mellin transforms, we find the leading  OPEs 
		\be
			\begin{split}
				G^+_{\Delta_1} (z_1, \bz_1) O^{+a}_{\Delta_2} (z_2, \bz_2) &\sim -\frac{ \kappa }{2}\frac{\bz_{12}}{z_{12}} B(\Delta_1-1, \Delta_2)O^{+a}_{\Delta_1 + \Delta_2}(z_2, \bz_2),\\
				G^+_{\Delta_1} (z_1, \bz_1) O^{-a}_{\Delta_2} (z_2, \bz_2)& \sim -\frac{\kappa}{2}\frac{\bz_{12}}{z_{12}} B(\Delta_1-1, \Delta_2 +2) O^{-a}_{\Delta_1 + \Delta_2}(z_2, \bz_2),
			\end{split}
		\ee
which agree with equation \eqref{rds}. 

Now we compute the graviton contribution to the mixed helicity gluon OPE. Since we are interested in the contribution from $G^-$ to the $O^+ O^-$ OPE, consider the on-shell vertex 
\be 
\begin{split}
\label{gg-G}
V(p_1, p_2, p_3) =- i \kappa \delta^{a_1 a_2}
						 \left[ (\ve_1^+ \cdot \ve_2^-) (\ve_3^+ \cdot p_1)(\ve_3^+ \cdot p_2)  
							- (\ve_1^+ \cdot p_2) (\ve_3^+ \cdot p_1)(\ve_2^{-} \cdot \ve_3^+)   \right].
\end{split}
\ee
Here $\varepsilon_1, \varepsilon_2$ are the polarizations of the positive and negative helicity gluons of momenta $p_1$, $p_2$ and colors $a_1$, $a_2$ respectively. $\varepsilon_{3\mu\nu}^{\pm} = \varepsilon_{3\mu}^{\pm} \varepsilon_{3\nu}^{\pm}$ is the graviton polarization.  Evaluating in our parametrization \eqref{mompar} and \eqref{pt}, the on-shell vertex  becomes
\be
	V(p_1, p_2, p_3) =- i \kappa \delta^{a_1 a_2} \omega_1 \omega_2 \bz_{13}^2.
\ee
In $(2,2)$ signature, the result is non-vanishing and upon taking $z_1 = z_2 = z_3$, momentum conservation reduces to 
\be
	\begin{split}
		\omega_1+\omega_2 + \omega_3 & =0,\\
			\omega_1 \bz_1+\omega_2 \bz_2+ \omega_3 \bz_3 & =0.
	\end{split}
\ee
Solving for $\bz_3$, we find
\be
	\bz_{3} = \frac{\omega_1}{\omega_1+ \omega_2} \bz_1 +  \frac{\omega_2}{\omega_1+ \omega_2} \bz_2   \quad \quad \quad \Rightarrow \quad \quad \quad \bz_{13}  =   \frac{\omega_2}{\omega_1+ \omega_2} \bz_{12}  .
\ee
Then, accounting for the graviton propagator,  we find that the collinear singularity for opposite helicity gluons due to the EYM vertex \eqref{gg-G} is
\be
{\rm Split}_{1-1}^{-2}(p_1,p_2) = \frac{\kappa}{2} \frac{\bz_{12}}{z_{12}} \frac{\omega_2^2}{\omega_P^2}.
\ee
Taking a Mellin transform, we deduce that the $O^{+a}_{\Delta_1}(z_1)O^{-b}_{\Delta_2}(z_2)$ OPE contains a term of the form
\be
\delta^{ab} \frac{\kappa}{2} \frac{\bz_{12}}{z_{12}} B(\Delta_1, \Delta_2 + 2) G^-_{\Delta_1 + \Delta_2}(z_2, \bz_2),
\ee
in agreement with the symmetry-derived result \eqref{H}. 

\section{Celestial incoming and outgoing OPEs}
\label{sec:in-out}

In this section we generalize our results to account for the presence of both incoming and outgoing particles. We introduce celestial operators 
\be
\mathcal{O}_{\Delta_k, s_k}^{\epsilon_k}(z_k, \bz_k) = \int_0^{\infty} d\omega_k~ \omega_k^{\Delta_k - 1} \mathcal{O}_{s_k}(\epsilon_k \omega_k, z_k, \bz_k)
\ee
carrying an additional label $\epsilon_k = \pm 1$ which distinguishes between outgoing and incoming states respectively. $\mathcal{O}_{s_k}(\epsilon_k \omega_k, z_k, \bz_k)$ are operators associated to
the standard `out' and `in' momentum eigenstates through the parametrization \eqref{mompar}. Since the action of the translation operator on `in' and `out' momentum eigenstates differs by a sign, the action of $\cp$ on the celestial operators
generalizes to 
\be \label{translation2}
	\delta_{\cp} \co^{ \eps}_{\Delta,s} (z,\bz) = \eps \co^{\eps}_{\Delta+1,s} (z, \bz).
\ee 
Note, since the `in' and `out' labels of asymptotic states are directly related to charges of the corresponding operators under a global symmetry of the celestial CFT, these labels are naturally a part of the celestial CFT data. 

Likewise, since the inverse of $\cp$ appears in the relevant subleading gluon and subsubleading graviton symmetry actions  (see appendices  \ref{ssga} and \ref{ssgta}), the actions of these symmetries 
\eqref{sgssfc} and \eqref{gssst} generalize to
\be
\label{sgssfcg}
\begin{split}
\bar{\delta}^a\mathcal{O}^{\epsilon_k}_{\Delta_k, s_k}(z_k, \bz_k) &=   \left[-\epsilon_k \left(\Delta_k - s_k - 1 +  \bz_k \p_{\bz_k} \right)T^a_k \mathcal{P}_k^{-1} - \frac{\kappa}{2} \bz_k \mathcal{F}_{k}^{+a} + \frac{\kappa}{2} \bz_k \mathcal{G}_k^{+a} \right]\mathcal{O}^{\epsilon_k}_{\Delta_k, s_k}(z_k, \bz_k),\\
\bar{\delta} \mathcal{O}^{\epsilon_k}_{\Delta_k, s_k} (z_k, \bz_k)&= -\frac{\kappa}{4}\epsilon_k \left[(\Delta_k - s_k)(\Delta_k - s_k - 1) +  4(\Delta_k - s_k) \bz_k \p_{\bz_k} + 3 \bz_k^2\p^2_{\bz_k}\right]\mathcal{O}_{\Delta_k-1, s_k}^{\epsilon_k} (z_k, \bz_k),
\end{split}
\ee
where $\mathcal{F}$ and $\mathcal{G}$ are defined in appendix \ref{ssga}.

\subsection{Gluon OPEs from asymptotic symmetries}
We now determine the OPE coefficients among outgoing and incoming gluons from \eqref{sgssfcg}. The case when both operators are incoming is mostly identical 
to the previously studied case with both operators outgoing since the symmetry constraints remain unchanged.  That is, up to  normalization, these OPE coefficients
are solved by  the Euler beta functions \eqref{solO+O+} and \eqref{solO+O-} for gluons of identical and opposite helicity respectively. We therefore consider the OPEs of outgoing and incoming gluons where as we will see, the 
constraints from symmetry differ.  

Generalizing \eqref{saz} and \eqref{sbz}, we begin with the ansatz
 \be \label{saz-in-out}
 \begin{split}
O^{+a, \epsilon}_{\D_1}(z_1,\bz_1) O^{+b, -\epsilon}_{\D_2}(z_2, \bz_2) \sim - \eps \frac{i f^{ab}_{~~c}}{z_{12}}& \left[  C' (\D_1, \D_2) O^{+c, \epsilon}_{\D_1 + \D_2-1}(z_2, \bz_2) \right. \\
&\left. + C''(\D_1, \D_2) O^{+c, -\epsilon}_{\D_1 + \D_2-1}(z_2, \bz_2) \right],
\end{split}
 \ee
 \be \label{sbz-in-out}
 \begin{split}
O^{+a, \epsilon}_{\D_1}(z_1,\bz_1) O^{-b, -\epsilon}_{\D_2}(z_2, \bz_2) \sim - \eps \frac{i f^{ab}_{~~c}}{z_{12}} &\left[ D'(\D_1, \D_2) O^{-c, \epsilon}_{\D_1 + \D_2-1}(z_2, \bz_2) \right. \\
&\left. + D''(\D_1, \D_2) O^{-c, -\epsilon}_{\D_1 + \D_2-1}(z_2, \bz_2) \right].
\end{split}
 \ee 
Using the generalized action of the translation operator \eqref{translation2}, we find the OPE coefficients must obey 
 \be \label{newt1}
 	\begin{split}
		C'(\Delta_1+1, \Delta_2) -C'(\Delta_1, \Delta_2+1)  &= C'(\Delta_1, \Delta_2),\\
		C''(\Delta_1+1, \Delta_2) -C''(\Delta_1, \Delta_2+1)  &=- C''(\Delta_1, \Delta_2),
	\end{split}
 \ee
 and
  \be\label{newt2}
 	\begin{split}
		D'(\Delta_1+1, \Delta_2) -D'(\Delta_1, \Delta_2+1)  &= D'(\Delta_1, \Delta_2),\\
		D''(\Delta_1+1, \Delta_2) -D''(\Delta_1, \Delta_2+1)  &=- D''(\Delta_1, \Delta_2).
	\end{split}
 \ee
 As before,  these recursion relations do not fully constrain the answer, so we turn to the subleading soft gluon symmetry.
Constraining \eqref{saz-in-out}  with the symmetry in \eqref{sgssfcg} and following the logic in section \ref{sec:3.1}, we obtain the following relations
\be 
\begin{split}
(\D_1-2)C'(\D_1-1,\D_2) f^a_{~dc}f^{cb}_{~~e} - (\D_2-2) C'(\D_1,\D_2-1)f^b_{~dc}f^{ac}_{~~e} \\ = (\D_1+\D_2-3)C'(\D_1, \D_2) f^{ab}_{~~c}f^c_{~de},\\
(\D_1-2)C''(\D_1-1,\D_2) f^a_{~dc}f^{cb}_{~~e} - (\D_2-2) C''(\D_1,\D_2-1)f^b_{~dc}f^{ac}_{~~e} \\ =- (\D_1+\D_2-3)C''(\D_1, \D_2) f^{ab}_{~~c}f^c_{~de},
\end{split}
 \ee
which using the Jacobi identity reduce to 
\be
\label{cout}
\begin{split}
(\Delta_1 - 2)C'(\Delta_1 - 1, \Delta_2) &= (\Delta_1 + \Delta_2 - 3) C'(\Delta_1, \Delta_2), \\
 -(\Delta_2 - 2)C'(\Delta_1, \Delta_2 - 1)
& = (\Delta_1 + \Delta_2 - 3)C'(\Delta_1, \Delta_2),
 \end{split}
\ee
and 
\be
\label{cin}
\begin{split}
(\Delta_1 - 2)C''(\Delta_1 - 1, \Delta_2)& = -(\Delta_1 + \Delta_2 - 3) C''(\Delta_1, \Delta_2), \\
 (\Delta_2 - 2)C''(\Delta_1, \Delta_2 - 1)
& = (\Delta_1 + \Delta_2 - 3)C''(\Delta_1, \Delta_2). 
 \end{split}
\ee
By shifting the arguments and taking a linear combination of the two constraints for each OPE coefficient, one can verify that these new recursion relations imply the modified recursion relation \eqref{newt1} from translation symmetry.
\eqref{cout} and \eqref{cin} are solved by 
\be 
\label{cec-e}
	\begin{split}
		C'(\Delta_1, \Delta_2) &=-   B(\Delta_2 - 1, 3 - \Delta_1 - \Delta_2),\\
		 C''(\Delta_1, \Delta_2)&=   B(\Delta_1 - 1, 3 - \Delta_1 - \Delta_2),
	\end{split}
\ee
where we have used the celestial soft gluon theorem, generalized for incoming and outgoing operators,
\be  
\label{clst}
\begin{split}
\lim_{\Delta_1 \rightarrow 1}O^{+a, \epsilon}_{\D_1}(z_1,\bz_1) O^{+b, -\epsilon}_{\D_2}(z_2, \bz_2) = -\epsilon \frac{if^{ab}{}_{c}}{z_{12}}\frac{1}{\Delta_1 - 1}O^{+c, -\epsilon}_{\D_2}(z_2, \bz_2) 
\end{split}
\ee
to fix the normalization. Note that both $C'$ and $C''$ are fixed by \eqref{clst} due to the symmetry of \eqref{saz-in-out} under exchange of labels which implies that they have soft poles at $\Delta_2, \Delta_1 = 1$ respectively, as seen explicitly in \eqref{cec-e}.    As we will see now, this will not usually be the case and a more general argument will be needed.

For opposite helicity gluons,  the soft gluon symmetry constraints on \eqref{sbz-in-out} reduce to 
\be \label{recglue1}
\begin{split}
(\Delta_1 - 2)D'(\Delta_1 - 1, \Delta_2) = (\Delta_1 + \Delta_2 - 1) D'(\Delta_1, \Delta_2), \\
 - \Delta_2 D'(\Delta_1, \Delta_2 - 1)
 = (\Delta_1 + \Delta_2 - 1)D'(\Delta_1, \Delta_2),
 \end{split}
\ee
\be \label{recglue2}
\begin{split}
(\Delta_1 - 2)D''(\Delta_1 - 1, \Delta_2) = -(\Delta_1 + \Delta_2 - 1) D''(\Delta_1, \Delta_2), \\
 \Delta_2 D''(\Delta_1, \Delta_2 - 1)
 = (\Delta_1 + \Delta_2 - 1)D''(\Delta_1, \Delta_2).
 \end{split}
\ee
The leading soft gluon theorem implies
\be  
\begin{split} 
\lim_{\Delta_1 \rightarrow 1}O^{+a, \epsilon}_{\D_1}(z_1,\bz_1) O^{-b, -\epsilon}_{\D_2}(z_2, \bz_2) = -\epsilon \frac{if^{ab}_{\ \ c}}{z_{12}}\frac{1}{\Delta_1 - 1}O^{- c, -\epsilon}_{\D_2}(z_2, \bz_2),
\end{split}
\ee
which together with the recursion relation \eqref{recglue2} uniquely fixes
\be
	D''(\Delta_1 , \Delta_2) =    B(\Delta_1 - 1,1 - \Delta_1 - \Delta_2).
\ee
On the other hand, \eqref{recglue1} is solved by 
\be
	D'(\Delta_1, \Delta_2) = \alpha B(\Delta_2 + 1, 1 - \Delta_1 - \Delta_2)
\ee
for some yet-to-be determined constant $\alpha$.  To fix $\alpha$, consider the mixed-helicity gluon OPE, evaluated at $\Delta_1 = \Delta_2 \equiv \Delta$
	\be 
					\begin{split}
						O^{+a, \eps}_{\Delta }& (z_1, \bz_1)O^{-b, -\eps}_{\Delta} (z_2, \bz_2)\\
							 &\sim \frac{ -i  f^{ab}{}_c }{z_{12}} \eps  \left[\alpha B\big( \Delta+1   , 1- 2\Delta   \big)  O^{-c, \eps}_{2\Delta  -1 }(z_2, \bz_2)
							 	+   B\big(\Delta - 1, 1-2\Delta  \big)O^{-c,-\eps}_{2\Delta  -1 }(z_2, \bz_2)\right] .
					\end{split}
			\ee
	Taking      $\Delta \to 1$, which corresponds to  a double soft limit of a scattering amplitude, we obtain an OPE among celestially soft operators
	\be 
					\begin{split}
						\lim_{\Delta \to 1} (\Delta -1)^2&O^{+a, \eps}_{\Delta } (z_1, \bz_1)O^{-b, -\eps}_{\Delta} (z_2, \bz_2)\\
							 &\sim \frac{ -i  f^{ab}{}_c }{z_{12}}\eps \frac{1}{2}  \lim_{\Delta \to 1}  \left[\alpha (\Delta -1)   O^{-c, \eps}_{2\Delta  -1 }(z_2, \bz_2)
							 	+  (\Delta -1) O^{-c,-\eps}_{2\Delta  -1 }(z_2, \bz_2)\right] .
					\end{split}
	\ee
	The above OPE is related to another OPE for celestially soft operators
	\be
		\begin{split}
			\lim_{\Delta \to 1} (\Delta -1)^2&O^{+a, \eps}_{\Delta } (z_1, \bz_1)O^{-b, \eps}_{\Delta} (z_2, \bz_2)
				 \sim -i f^{ab}{}_c \frac{\eps}{z_{12}} \lim_{\Delta \to 1} (\Delta - 1)O^{-c, \eps}_{2\Delta-1 } (z_2, \bz_2) 
		\end{split}
	\ee
	by  the crossing relation for soft modes \cite{He:2015zea}, which on the celestial sphere takes the form
	\be
		\lim_{\Delta \to 1} (\Delta-1) O^{\pm a, \eps}_{\Delta}(z, \bz) = -\lim_{\Delta \to 1} (\Delta-1) O^{\pm a,- \eps}_{\Delta}(z, \bz) .
	\ee
	Comparing the two, we find
	\be
				 \alpha= -1.
			\ee  
	  \subsection{Gluon OPEs from collinear singularities}
We now confirm the symmetry-derived results  from a momentum-space amplitude calculation.  
As before, the OPE coefficients can be derived by Mellin transforming the collinear splitting functions. For incoming and outgoing gluons these take the general form
\be
{\rm Split}_{1 s_2}^{s_2} (p_1 , p_2)= \frac{1}{z_{12}}(\epsilon_1 \omega_1)^{\alpha}(\epsilon_2\omega_2)^{\beta}(\epsilon_1\omega_1 + \epsilon_2\omega_2)^{\gamma} .
\ee
To evaluate
\be 
\label{mella}
\int_0^{\infty} d\omega_1 \int_0^{\infty} d\omega_2~ \omega_1^{\Delta_1 - 1} \omega_2^{\Delta_2 - 1} {\rm Split}^{s_2}_{1 s_2}  (p_1 , p_2),
\ee
it is convenient to make the following change of variables
\be
\omega_1 = (1 - \epsilon_1\epsilon_2 t)\omega_P, \qquad \omega_2 = t \omega_P,
\ee
where
\be
\omega_1 + \epsilon_1 \epsilon_2 \omega_2 = \omega_P. 
\ee
For $\epsilon_1 \epsilon_2 = -1$, \eqref{mella} splits into two integrals such that the celestial OPE takes the form 
\be
	\begin{split}
	O^{+a, \eps_1}_{\Delta_1} (z_1, \bz_1)
			O^{\pm b, \eps_2}_{\Delta_2}& (z_2, \bz_2)\\
				   \sim \frac{-i f^{ab}{}_c}{z_{12}} \eps_1^{\alpha + \gamma} \eps_2^\beta 
					&\left[\int_0^{\infty} d\omega_P \int_0^{\infty} dt~ (1+t)^{\Delta_1 - 1 + \alpha} t^{\Delta_2 - 1 + \beta} \omega_P^{\Delta_1 + \Delta_2 + \alpha + \beta + \gamma - 1}  O^{\pm c} (\eps_1 \omega_P, z_2, \bz_2)\right. \\
					&\left.-\int^0_{-\infty} d\omega_P \int^{-1}_{-\infty} dt ~(1+t)^{\Delta_1 - 1 + \alpha} t^{\Delta_2 - 1 + \beta} \omega_P^{\Delta_1 + \Delta_2 + \alpha + \beta + \gamma - 1}
						O^{\pm c} (\eps_1 \omega_P, z_2, \bz_2)\right]\\
				= \frac{-i f^{ab}{}_c}{z_{12}} \eps_1^{\alpha + \gamma} \eps_2^\beta 
					&\left[  \int_0^{\infty} dt~ (1+t)^{\Delta_1 - 1 + \alpha} t^{\Delta_2 - 1 + \beta}  
							O^{\pm c, \eps_1}_{\Delta_1 + \Delta_2 + \alpha + \beta + \gamma} (  z_2, \bz_2)\right. \\
					&\left.+ (-1)^\gamma   \int_{0}^{\infty} dt ~t^{\Delta_1 - 1 + \alpha}(1+ t)^{\Delta_2 - 1 + \beta} O^{\pm c,-\eps_1}_{\Delta_1 + \Delta_2 + \alpha + \beta + \gamma} (  z_2, \bz_2)\right],
	\end{split}
\ee
where to obtain the second line, we performed the change of variables $\omega_P \rightarrow -\omega_P$ and  $ t \to -(1+t)$ on the second term.   Upon making a further change of variables
$t = \dfrac{u}{1-u}$, we find the remaining $t$-integrals once again take the form \eqref{betafunction} so that the OPE coefficients are given by Euler beta functions
\be
	\begin{split}
	O^{+a, \eps_1}_{\Delta_1} (z_1, \bz_1)
			O^{\pm b, \eps_2}_{\Delta_2}& (z_2, \bz_2)\\
				   \sim  \frac{-i f^{ab}{}_c}{z_{12}} \eps_1^{\alpha + \gamma} \eps_2^\beta 
					&\left[ B(\Delta_2+ \beta, 1- \Delta_1-\Delta_2-\alpha-\beta)  O^{\pm c, \eps_1}_{\Delta_1 + \Delta_2 + \alpha + \beta + \gamma} (  z_2, \bz_2)\right. \\
					&\left.+ (-1)^\gamma  B(\Delta_1+ \alpha, 1- \Delta_1-\Delta_2-\alpha-\beta)  O^{\pm c,-\eps_1}_{\Delta_1 + \Delta_2 + \alpha + \beta + \gamma} (  z_2, \bz_2)\right].
	\end{split}
\ee 
For equal helicity gluons $\alpha = \beta = -\gamma = -1$ and so the in/out OPE is
\be 
\begin{split}
O^{+ a, \epsilon}_{\Delta_1}(z_1,\bz_1)O^{ + b, -\epsilon}_{\Delta_2}(z_2, \bz_2) \sim i f^{ab}_{\ \ c}\frac{\epsilon}{z_{12}}\left[ B(3 -\Delta_1 - \Delta_2 , \Delta_2 -1 )O_{\Delta_1 + \Delta_2 -1}^{+c,\epsilon}(z_2,\bz_2)\right.\\
\left. - B(\Delta_1 -1, 3 - \Delta_1 - \Delta_2 )O_{\Delta_1 + \Delta_2 -1}^{+c, -\epsilon}(z_2, \bz_2)\right]. 
\end{split}
\ee
For opposite helicity gluons $\alpha = -\beta = \gamma  = -1$ and we find
\be 
\begin{split}
O^{+ a, \epsilon}_{\Delta_1}(z_1, \bz_1)O^{ - b, -\epsilon}_{\Delta_2}(z_2, \bz_2) \sim i f^{ab}_{\ \ c}\frac{\epsilon}{z_{12}}\left[ B( -\Delta_1 - \Delta_2 + 1, \Delta_2 + 1 )O_{\Delta_1 + \Delta_2 -1}^{-c,\epsilon}(z_2, \bz_2)\right.\\
\left. -  B(\Delta_1 -1, 1 - \Delta_1 - \Delta_2 )O_{\Delta_1 + \Delta_2 -1}^{-c, -\epsilon}(z_2, \bz_2)\right],
\end{split}
\ee
which agree with the symmetry-derived OPEs. Analogous computations yield the graviton and gluon-graviton in/out OPEs. We summarize the results in the following section. 

\subsection{Summary of OPE coefficients}
\label{sec:3.4}
In summary, all the nonzero leading $z_{12}$ poles for all possible configurations of incoming and outgoing gluon and graviton OPEs are determined by the asymptotic symmetries in tree-level EYM.
The equal helicity gluon OPEs are
\be  
		\begin{split}
			O_{\Delta_1}^{+a, \eps} ( z_1, \bz_1)
				 O_{\Delta_2}^{+b, \eps}  (  z_2, \bz_2) 
					& \sim \frac{ -i  f^{ab}{}_c }{z_{12} } \eps
					B(\Delta_1-1, \Delta_2-1)     O_{\Delta_1 +\Delta_2 -1}^{+c ,\eps} (  z_2, \bz_2) ,
				\\
			  O_{\Delta_1}^{+a,\eps} ( z_1, \bz_1)
				 O_{\Delta_2}^{+b,-\eps}  (  z_2, \bz_2) 
					 & \sim \frac{ -i  f^{ab}{}_c}{z_{12}} \eps
					\left[ -   B( \Delta_2-1,3- \Delta_1- \Delta_2 )       O_{\Delta_1 +\Delta_2 -1}^{+c,\eps} (  z_2, \bz_2)\right.\\& \quad \quad \quad \quad \quad \quad 
				\left.	+B(\Delta_1-1,3- \Delta_1- \Delta_2 )    O_{\Delta_1 +\Delta_2 -1}^{+c,-\eps} (  z_2, \bz_2)
						\right].
						\end{split}
						\ee
						The mixed helicity gluon OPEs are
						\be
						\begin{split}
			O_{\Delta_1}^{+a,\eps} ( z_1, \bz_1)
				 O_{\Delta_2}^{-b,\eps}  (  z_2, \bz_2) 
					&  \sim \frac{ -i  f^{ab}{}_c }{z_{12} }  \eps
					B(\Delta_1-1, \Delta_2+ 1)     O_{\Delta_1 +\Delta_2-1}^{ -c, \eps} (  z_2, \bz_2)\\& \quad \quad 
					+  \frac{\kappa}{2} \frac{\bz_{12}}{z_{12}} \delta^{ab}
					B(\Delta_1 , \Delta_2+2)     G_{\Delta_1 +\Delta_2  }^{-,\eps} (  z_2, \bz_2),
\\
			  O_{\Delta_1}^{+a,\eps} ( z_1, \bz_1)
				 O_{\Delta_2}^{-b,-\eps}  (  z_2, \bz_2) 
					& \sim \frac{ -i  f^{ab}{}_c   }{z_{12}}\eps
					\left[ -   B( \Delta_2+ 1,1- \Delta_1- \Delta_2 )      O_{\Delta_1 +\Delta_2 -1}^{-c, \eps} (  z_2, \bz_2)\right.\\& \quad \quad \quad \quad \quad  \quad 
					 \left.+ B(\Delta_1-1,1- \Delta_1- \Delta_2 )   O_{\Delta_1 +\Delta_2 -1}^{-c,-\eps} (  z_2, \bz_2)\right]\\& \quad \quad +
						  \frac{\kappa}{2}\frac{\bz_{12}}{z_{12}}\delta^{ab}
					\left[  B( \Delta_2+2,-1- \Delta_1- \Delta_2 )  
						    G_{\Delta_1 +\Delta_2  }^{-,\eps} (  z_2, \bz_2)\right. \\& \quad \quad \quad  \quad  \quad  \quad  \quad 
				\left.	+   B(\Delta_1 ,-1- \Delta_1- \Delta_2 )  G_{\Delta_1 +\Delta_2  }^{-,-\eps} (  z_2, \bz_2)\right].
		\end{split}
	\ee  
The graviton OPEs are
	\be  
		\begin{split}
			G_{\Delta_1}^{+,\eps} ( z_1, \bz_1)
				 G_{\Delta_2}^{\pm, \eps}  (  z_2, \bz_2) 
					&  \sim - \frac{\kappa}{2}  \frac{\bz_{12}}{z_{12} }   
					B(\Delta_1-1, \Delta_2+1 \mp 2)     G_{\Delta_1 +\Delta_2 }^{\pm, \eps} (  z_2, \bz_2),
\\
			  G_{\Delta_1}^{+,\eps} ( z_1, \bz_1)
				 G_{\Delta_2}^{\pm ,-\eps}  (  z_2, \bz_2) 
					 & \sim  \frac{\kappa}{2}  \frac{\bz_{12}}{z_{12}}
					\left[    B( \Delta_2+1 \mp 2,1 \pm 2- \Delta_1- \Delta_2 )      G_{\Delta_1 +\Delta_2 }^{\pm,\eps} (  z_2, \bz_2)\right.\\&  \quad \quad \quad \quad 
					\left.+B(\Delta_1-1,1 \pm 2- \Delta_1- \Delta_2 )  G_{\Delta_1 +\Delta_2 }^{\pm,-\eps} (  z_2, \bz_2) \right].
		\end{split}
	\ee 
	The gluon-graviton OPEs are
 	\be   
		\begin{split}
		G_{\Delta_1}^{+,\eps} ( z_1, \bz_1)
				 O_{\Delta_2}^{\pm a,\eps}  (  z_2, \bz_2) 
					&  \sim - \frac{\kappa}{2} \frac{\bz_{12}}{z_{12}} 
					B(\Delta_1-1, \Delta_2+1 \mp 1)     O_{\Delta_1 +\Delta_2  }^{\pm a,\eps} (  z_2, \bz_2),
\\
			  G_{\Delta_1}^{+,\eps} ( z_1, \bz_1)
				 O_{\Delta_2}^{\pm a,-\eps}  (  z_2, \bz_2) 
					& \sim - \frac{\kappa}{2} \frac{\bz_{12}}{z_{12}}
					\left[   B( \Delta_2+1 \mp 1,1 \pm 1- \Delta_1- \Delta_2 )  
						  O_{\Delta_1 +\Delta_2  }^{\pm a,\eps} (  z_2, \bz_2)\right. \\& \quad \quad \quad \quad \quad 
					\left.-  B(\Delta_1-1,1 \pm 1- \Delta_1- \Delta_2 )    O_{\Delta_1 +\Delta_2  }^{\pm a,-\eps} (  z_2, \bz_2)\right].
		\end{split}
	\ee 
	From \eqref{gen_norm}, we recall a factor of $g_{YM}$ is absorbed in $f^{ab}_{\ \ c}$. The $\bz_{12} \rightarrow 0$ celestial OPEs are obtained in a similar way by imposing the $\delta$ symmetry instead. 
 	 
 The presence of  higher-dimension operators due to quantum, stringy or other corrections is  expected to augment this list with the finite number of additions  allowed by the general formula \eqref{gfms}. A list of all possible corrections in theories with only gluons and gravitons is given in appendix \ref{ft}.

\section*{Acknowledgements}
We are   grateful  to Nima Arkani-Hamed, Prahar Mitra,  Sabrina Pasterski, Andrea Puhm, Shu-Heng Shao, Mark Spradlin  and Tom Taylor  for useful conversations. This work was supported by NSF  grant  1205550 and the John Templeton Foundation. M.P. acknowledges the support of a Junior Fellowship at the Harvard Society of Fellows. 

\appendix

\section{Celestial OPEs from bulk three-point vertices}
\label{celOPE}

In this appendix we relate the conformal weights of the operators which are allowed to appear in the OPE of two conformal primaries to the bulk dimensions of the corresponding three-point vertices.
We consider a bulk three-point vertex among gluons and gravitons which schematically takes the form
\be
\label{vertex}
V = \p^{m} \Phi_1(x) \Phi_2(x)\Phi_3(x) ,
\ee
where $\Phi_1, \Phi_2, \Phi_3 \in \{{A_{\mu}, h_{\mu\nu}}\}$,  and we omitted Lorentz indices which should be contracted accordingly. $m$ is the total number of derivatives in the interaction, which are appropriately distributed among $\Phi_1, \Phi_2, \Phi_3$. Since both gluons and gravitons have dimension $1$, the net dimension of the vertex is 
\be
\label{dv}
d_V = 3 + m. 
\ee 

Suppose $\Phi_1, \Phi_2$ are taken to be outgoing external legs (on-shell states). In momentum space, each derivative is associated with a factor of momentum. Upon parametrizing momenta as in \eqref{mompar}, Mellin transforming with respect to $\omega_1$ and $\omega_2$ and taking the collinear limit $z_{12} \rightarrow 0$, the celestial amplitude takes the general form
\be
\widetilde{\mathcal{A}} = \sum_{\alpha, \beta} \int_0^{\infty} d\omega_1 \int_0^{\infty} d\omega_2 ~\omega_1^{\Delta_1 - 1}\omega_2^{\Delta_2 - 1} \frac{\omega_1^{m + \alpha} \omega_2^{\beta }}{\omega_P^{\alpha + \beta}} \frac{1}{\omega_1 \omega_2} F_{\alpha, \beta}(z_1, \bz_1, z_2, \bz_2; \cdots),
\ee
where we used momentum conservation and  accounted for the $\Phi_3$ propagator.  Since we're working in a collinear expansion, $F_{\alpha, \beta}$ depends only on $\omega_P$, but not $\omega_1$ or $\omega_2$ independently.  
In general, the amplitude involves a sum over terms with different $\alpha, \beta$. The details depend on the precise form of the interaction but turn out to be irrelevant in determining the scaling dimension of the allowed operators. 
Setting
\be
  \omega_1 = \omega_P t, \quad \quad \quad \omega_2 = \omega_P (1-t),
\ee 
the celestial amplitude becomes
\be 
\widetilde{\mathcal{A}} = \sum_{\alpha, \beta} B(\Delta_1 + m + \alpha - 1, \Delta_2 + \beta - 1) \int_0^{\infty} d\omega_P\omega_P^{ \Delta_1 + \Delta_2 - 3 +m} F_{\alpha, \beta}(z_1, \bz_1, z_2, \bz_2; \omega_P, \cdots ).
\ee
This allows one to read off the scaling dimension of the associated operator in the OPE expansion
\be
\Delta_3 - 1 =  \Delta_1 + \Delta_2 - 3 + m \implies \Delta_3 = \Delta_1 + \Delta_2 + d_V - 5,
\ee
where in the last equation we used \eqref{dv}. We therefore conclude that the primaries in the $\Phi_1, \Phi_2$ OPE can be classified according to the dimension of the possible corresponding bulk three-point vertices as in \eqref{gfms}. 

\section{Higher order  OPEs}
\label{ft}

There is a finite number of primaries which contribute to the OPE \eqref{gfms} to any finite order in the  $z_{12}$ expansion. To get a flavor of this, in this appendix  we collect all possible single-pole or finite terms. 
For  the gluon-gluon OPEs these are
\be
		\begin{split}
			O^{+a}_{\Delta_1} (z_1, \bz_1)O^{+b}_{\Delta_2} (z_2, \bz_2):\  &\frac{1}{z_{12}} O^{+c}_{\Delta_1+\Delta_2 -1}(z_2, \bz_2), \ \ 
										\frac{\bz_{12}^2}{z_{12}} O^{-c}_{\Delta_1+\Delta_2 +1}(z_2, \bz_2),\\
										&\bz_{12}  O^{+c}_{\Delta_1+\Delta_2 +1}(z_2, \bz_2),\ \  \bz_{12}^3  O^{-c}_{\Delta_1+\Delta_2 +3}(z_2, \bz_2),\\
										& G^+_{\Delta_1 +\Delta_2 } (z_2, \bz_2),\ \ \frac{\bz_{12}^3}{z_{12}}  G^-_{\Delta_1 +\Delta_2+2 } (z_2, \bz_2), \ \
										\bz_{12}^4 G^-_{\Delta_1 +\Delta_2+4 } (z_2, \bz_2),\\
			O^{+a}_{\Delta_1} (z_1, \bz_1)O^{-b}_{\Delta_2} (z_2, \bz_2):\  &
		 \frac{1}{z_{12}}  O^{-c}_{\Delta_1+\Delta_2 -1}(z_2, \bz_2), \ \ \bz_{12}  O^{-c}_{\Delta_1+\Delta_2 +1}(z_2, \bz_2),\\
		 & \frac{\bz_{12}}{z_{12}}  G^-_{\Delta_1 +\Delta_2 } (z_2, \bz_2),\ \   \bz_{12}^2    G^-_{\Delta_1 +\Delta_2+2 } (z_2, \bz_2) .
		\end{split}
	\ee	Operators on the right hand side of dimension $\Delta_1 + \Delta_2 - 1$  arise from the pure YM three-point vertices while those of dimension $\Delta_1+ \Delta_2$ from  three-point vertices in EYM (excluding the three-gluon vertex).  All other operators of dimensions $\Delta_1 + \Delta_2 + n,\  n  = 1,...,4$ correspond to the following higher derivative vertices in order: $F^3, RF^2, \p^2 F^3, \p^2 R F^2.$
	
Similarily, the finite or single pole terms in the graviton-graviton OPE are
	\be
		\begin{split}
			G^+_{\Delta_1}(z_1, \bz_1)G^+_{\Delta_2}(z_2, \bz_2):\  &\frac{\bz_{12}}{z_{12}}  G^+_{\Delta_1+ \Delta_2}(z_2, \bz_2), \ \ \bz_{12}^2 G^+_{\Delta_1+ \Delta_2+2}(z_2, \bz_2), \\
					 & \frac{\bz_{12}^5}{z_{12}} G^-_{\Delta_1+ \Delta_2+4}(z_2, \bz_2), \ \  \bz_{12}^6  G^-_{\Delta_1+ \Delta_2+6}(z_2, \bz_2),\\
			G^+_{\Delta_1}(z_1, \bz_1)G^-_{\Delta_2}(z_2, \bz_2):\ &\frac{\bz_{12}}{z_{12}} G^-_{\Delta_1+ \Delta_2}(z_2, \bz_2),\ \   \bz_{12}^2  G^-_{\Delta_1+ \Delta_2+2}(z_2, \bz_2) .
		\end{split}
	\ee
	Operators of dimensions $\Delta_1 + \Delta_2 + n,\  n  = 2,4,6$ arise from the following higher derivative vertices in order: $R^2, R^3,  \p^2 R^3.$ The coefficient of the $R^2$ term can be eliminated by field redefinition\cite{tHooft:74}. 
 	
	The finite or single pole terms  in the gluon-graviton OPEs are
	\be
		\begin{split}
			G^+_{\Delta_1}(z_1, \bz_1)O^{+a}_{\Delta_2}(z_2, \bz_2):\  &\frac{\bz_{12}}{z_{12}} O^{+a}_{\Delta_1+ \Delta_2}(z_2, \bz_2),\ \  \frac{\bz_{12}^3}{z_{12}}O^{-a}_{\Delta_1+ \Delta_2+2}(z_2, \bz_2),\\
					  &\bz_{12}^2 O^{+a}_{\Delta_1+ \Delta_2+2}(z_2, \bz_2), \ \ \bz_{12}^4 O^{-a}_{\Delta_1+ \Delta_2+4}(z_2, \bz_2),\\
			G^+_{\Delta_1}(z_1, \bz_1)O^{-a}_{\Delta_2}(z_2, \bz_2):\  &\frac{\bz_{12}}{z_{12}}  O^{-a}_{\Delta_1+ \Delta_2}(z_2, \bz_2),\ \ 
			O^{+a}_{\Delta_1+ \Delta_2}(z_2, \bz_2),\ \   
			\bz_{12}^2 O^{-a}_{\Delta_1+ \Delta_2+2}(z_2, \bz_2).
		\end{split}
	\ee  
Operators of dimensions $\Delta_1 + \Delta_2 + n,\  n  = 2,4$ correspond to the higher derivative vertices $R F^2$ and $ \p^2 R F^2$ respectively.

\section{Subleading soft gluon symmetry}
\label{ssga}

Tree-level gauge theory amplitudes obey the  soft relation  (see \cite{Lysov:2014csa} and references therein)
\be
 \mathcal{A}^a_{n+1}(p_1,...,p_n;q) = \left( J^{a}_{(0)} + J^{a}_{(1)}  \right)\mathcal{A}_n(p_1,..., p_n) + \mathcal{O}(q),
\ee
where $J^{a}_{(0)}, J^{a}_{(1)}$ are the leading and subleading gluon soft factors and we suppressed all color indices except for $a$, the one associated with the soft gluon. 
In this section we derive the action of the subleading soft gluon symmetry on outgoing gluons  in a conformal basis. The subleading soft gluon operators are 
\be
\label{subl-sf}
J^{\pm a}_{(1)} =  \sum_{k = 1}^ni \frac{\varepsilon^{\pm}_{\mu} q_{\nu} \mathcal{J}^{\mu\nu}_k}{q\cdot p_k}T^a_k,
\ee
where $T^a_k$ are the generators of the non-abelian gauge group in representation $k$. In the parametrization \eqref{mompar} and \eqref{pt}, \eqref{subl-sf} takes the form
\be
\begin{split}
\label{sgssf}
 J^{-a}_{(1)}  &= \sum_{k = 1}^n \frac{1}{ \bz - \bz_k}\left(-\frac{s_k}{\omega_k} + \p_{\omega_k} + \frac{z - z_k}{\omega_k}\p_{z_k} \right)T^a_k, \\
 J^{+a}_{(1)} & =  \sum_{k = 1}^n \frac{1}{z - z_k}\left(\frac{s_k}{\omega_k} + \p_{\omega_k} + \frac{\bz - \bz_k}{\omega_k}\p_{\bz_k} \right)T^a_k.
 \end{split}
\ee
Upon performing a Mellin transform we find
\be 
\begin{split}
\label{sgssfMs}
 J^{-a}_{(1)} = \sum_{k = 1}^n \frac{1}{ \bz - \bz_k}\left(-2h_k + 1  + (z - z_k)\p_{z_k} \right)T^a_k  \mathcal{P}^{-1}_k , \\
 J^{+a}_{(1)}  =  \sum_{k = 1}^n \frac{1}{z - z_k}\left(-2\bh_k + 1 + (\bz - \bz_k)\p_{\bz_k} \right)T^a_k  \mathcal{P}^{-1}_k,
 \end{split}
\ee
where 
\be
h_k = \frac{1}{2}(\Delta_k+s_k), \qquad \bh_k = \frac{1}{2}( \Delta_k-s_k),
\ee
and $\mathcal{P}^{-1}_k$ implements the inverse shift on the $k^{\rm th}$ operator to the one defined in \eqref{Psym}.
Treating $z, \bz$ as independent complex variables, we can define the operators
\be 
\begin{split}
\delta^a \equiv \oint \frac{d\bz}{2\pi i} J^{-a}_{(1)}(0, \bz) = \lim_{\Delta \rightarrow 0}\Delta \oint \frac{d\bz}{2\pi i}{O}_{\Delta}^{- a}(0,\bz),\\ 
\bar{\delta}^a \equiv \oint \frac{dz}{2\pi i} J^{+a}_{(1)}(z, 0) = \lim_{\Delta \rightarrow 0}\Delta \oint \frac{dz}{2\pi i} {O}_{\Delta}^{+a}(z, 0),
\end{split}
\ee
which have the following action on gluons 
\be
\label{gsa}
\begin{split}
\delta_a O^{\pm b}_{\Delta_k}(z_k, \bz_k) =- if^b_{\ ac}\left(2h_k - 1  +  z_k \p_{z_k} \right) O^{\pm c}_{\Delta_k - 1}(z_k, \bz_k), \\
\bar{\delta}_a O^{\pm b}_{\Delta_k}(z_k, \bz_k) = -if^b_{\ ac}\left(2\bh_k  - 1  +  \bz_k \p_{\bz_k} \right) O^{\pm c}_{\Delta_k - 1}(z_k, \bz_k).
\end{split}
\ee
Equations \eqref{gsa} define a global symmetry associated with the subleading soft gluon theorem and constrain the gluon OPE coefficients as in \eqref{szz}.

$J_{(1)}$ receives corrections in the presence of gravitons. These can be deduced from the vertex \eqref{gg-G} in which case \eqref{sgssf} becomes
\be
\begin{split}
\label{sgssfc}
 J^{-a}_{(1)}  &= \sum_{k = 1}^n \frac{1}{ \bz - \bz_k}\left(-\frac{s_k}{\omega_k} + \p_{\omega_k} + \frac{z - z_k}{\omega_k}\p_{z_k} \right)T^a_k + \frac{\kappa}{2}\frac{z - z_k}{\bz - \bz_k}\mathcal{F}_{k}^{-a} - \frac{\kappa}{2}\frac{z - z_k}{\bz - \bz_k}\mathcal{G}_k^{-a}, \\
 J^{+a}_{(1)} & =  \sum_{k = 1}^n \frac{1}{z - z_k}\left(\frac{s_k}{\omega_k} + \p_{\omega_k} + \frac{\bz - \bz_k}{\omega_k}\p_{\bz_k} \right)T^a_k + \frac{\kappa}{2}\frac{\bz - \bz_k}{z - z_k}\mathcal{F}_{k}^{+a} - \frac{\kappa}{2}\frac{\bz -\bz_k}{z - z_k}\mathcal{G}_k^{+a},
 \end{split}
\ee
where
\be
	\begin{split}
		\mathcal{F}^{\pm a}_k&|p_k, s_k = \mp 1, a_k\rangle = \delta^{a a_k}|p_k, s_k = \mp 2\rangle, \quad \quad \quad 
		 \mathcal{F}^{\pm a}_k|p_k, s_k = \pm 1, a_k\rangle =0, \\
		 \mathcal{G}^{\pm a}_k&|p_k, s_k =  \pm 2 \rangle = \delta^{a a_k}|p_k, s_k =\pm 1, a_k\rangle, \quad \quad \quad 
	   \mathcal{G}^{\pm a}_k|p_k, s_k = \mp 2 \rangle = 0.
	\end{split}
\ee
This implies that \eqref{dsx} obeys 
\be
\lim_{\Delta_1 \rightarrow 0}\Delta_1 O_{\Delta_1}^{+a}(z_1,\bz_1) O^{-b}_{\Delta_2}(z_2,\bz_2) = \frac{i f^{ab}_{\ \ c}}{z_{12}} \Delta_2 O^{-c}_{ \Delta_2 - 1}(z_2,\bz_2) + \frac{\kappa}{2}\frac{\bz_{12}}{z_{12}} \delta^{ab} G^-_{ \Delta_2}(z_2,\bz_2),
\ee
which fixes the normalization of the graviton OPE coefficient.

\section{Subsubleading soft graviton symmetry}
\label{ssgta} 

In this section we derive the symmetry actions \eqref{sssc} and \eqref{sssggl} (for outgoing particles) from the subsubleading soft graviton theorem.

Tree-level gravity amplitudes were shown in \cite{Cachazo:2014fwa} to obey the following soft relation
\be 
\mathcal{A}_{n+1}(p_1,...,p_n;q) = \left( S_{(0)} + S_{(1)} + S_{(2)} \right)\mathcal{A}_n(p_1,..., p_n) + \mathcal{O}(q^2),
\ee
where $S_{(0)}, S_{(1)}$ and $S_{(2)}$ are the leading, subleading and subsubleading soft factors respectively. In this appendix we focus on the subsubleading soft factor, 
\be
\label{SF}
S_{(2)} = -\frac{\kappa}{4}\sum_{k = 1}^n\frac{\varepsilon_{\mu\nu} (q_{\rho}\mathcal{J}_k^{\rho\mu})(q_{\sigma}\mathcal{J}^{\sigma\nu}_k)}{q \cdot p_k}, 
\ee
where $\varepsilon_{\mu\nu}$ and $q$ are the polarization and momentum of the soft graviton and $\mathcal{J}_i, p_i$ are the total angular momenta and momenta of the hard particles. 
Using the parametrizations \eqref{mompar} of momenta and the angular momentum operators in \cite{Kapec:2016jld}, \eqref{SF} can be shown to reduce to \cite{Conde:2016rom,Guevara:2019ypd}
\be
\label{slsf}
\begin{split}
S_{(2)}^- = -\frac{\kappa}{4} \sum_{k = 1}^n\frac{\omega}{\omega_k} \frac{1}{(z - z_k)(\bz - \bz_k)}\left[(z - z_k)(s_k - \omega_k\p_{\omega_k}) - (z - z_k)^2\p_{z_k} \right]^2,\\
S_{(2)}^+ = -\frac{\kappa}{4} \sum_{k = 1}^n \frac{\omega}{\omega_k}\frac{1}{(z - z_k)(\bz - \bz_k)}\left[(\bz - \bz_k)(-s_k - \omega_k\p_{\omega_k}) - (\bz - \bz_k)^2\p_{\bz_k}\right]^2
\end{split}
\ee
for negative and positive helicity soft gravitons respectively.
In a conformal basis, \eqref{slsf} become
\be
\label{mellsfc}
\begin{split}
 \widetilde{S}_{(2)}^- = -\frac{\kappa}{4} \sum_{k = 1}^n \frac{z - z_k}{\bz - \bz_k}\left[2h_k(2h_k - 1) -  2(z - z_k) 2h_k \p_{z_k} + (z - z_k)^2\p^2_{z_k}\right]\mathcal{P}^{-1}_k ,\\ 
\widetilde{S}_{(2)}^+ = -\frac{\kappa}{4} \sum_{k = 1}^n \frac{\bz - \bz_k}{z - z_k}\left[2\bar{h}_k(2\bar{h}_k - 1)-  2(\bz - \bz_k) 2\bh_k \p_{\bz_k} + (\bz - \bz_k)^2\p^2_{\bz_k}\right]\mathcal{P}^{-1}_k,
\end{split}
\ee
with $h_k, \bh_k$ and $\mathcal{P}^{-1}$ defined in appendix \ref{ssga}. Treating $z, \bz$ as independent complex variables, we define the soft operators 
\be 
\begin{split}
\delta \equiv \oint \frac{d\bz}{2\pi i} \p_z \widetilde{S}_{(2)}^-(0, \bz) &= \lim_{\Delta \rightarrow -1} (\Delta + 1)\oint \frac{d\bz}{2\pi i} \p_z G_{\Delta}^-(0, \bz),\\
\bar{\delta} \equiv \oint \frac{dz}{2\pi i} \p_{\bz} \widetilde{S}_{(2)}^+(z, 0) &= \lim_{\Delta \rightarrow -1} (\Delta + 1)\oint \frac{dz}{2\pi i} \p_{\bz} G_{\Delta}^+(z, 0),
\end{split}
\ee
which act on celestial operators as follows
\be 
\label{gssst}
\begin{split}
\delta \mathcal{O}_{\Delta_k, s_k} (z_k, \bz_k) &= -\frac{\kappa}{4}\left[2h_k(2h_k - 1) +  8h_k  z_k \p_{z_k} + 3 z_k^2\p^2_{z_k}\right]\mathcal{O}_{\Delta_k - 1, s_k}(z_k, \bz_k),\\
\bar{\delta} \mathcal{O}_{\Delta_k, s_k} (z_k, \bz_k)&= -\frac{\kappa}{4}\left[2\bh_k(2\bh_k - 1) +  8\bh_k  \bz_k \p_{\bz_k} + 3 \bz_k^2\p^2_{\bz_k}\right]\mathcal{O}_{\Delta_k-1, s_k} (z_k, \bz_k).
\end{split}
\ee
Equations \eqref{gssst} define the action of the global symmetries associated with the subsubleading soft graviton theorem \eqref{sssc} and \eqref{sssggl}. They constrain the form of the graviton and graviton-gluon OPEs \eqref{gsaz} and \eqref{gsaz1} as discussed in sections \ref{g} and \ref{gg}.

\section{Solving the recursion relations}
\label{beta}

Consider a symmetric function of complex variables $C(\Delta_1, \Delta_2) = C(\Delta_2, \Delta_1)$ which obeys the recursion relation
\be
\label{rr}
\Delta_1 C(\Delta_1, \Delta_2) = (\Delta_1 + \Delta_2) C(\Delta_1 + 1, \Delta_2) .
\ee
Provided $C(\Delta_1, \Delta_2) \Gamma(\Delta_1 +\Delta_2)$ is holomorphic for ${\rm Re} (\Delta_1 )> 0 $ and bounded for ${\rm Re}( \Delta_1) \in [1, 2)$, \eqref{rr} has the unique solution
\be
C(\Delta_1, \Delta_2) =  C(1, 1) B(\Delta_1, \Delta_2).  
\ee
This can be proven in the following way. Define a function $f(x) \equiv C(x, y_0) \Gamma(x + y_0)$. Then \eqref{rr} becomes 
\be
\label{rr1}
x f(x) = f (x+1). 
\ee
By Wieland's theorem \cite{reichfunctional}, \eqref{rr1} has the unique solution
\be
f(x) = f(1) \Gamma(x).
\ee
Eliminating $f(x), f(1)$ in terms of $C(x, y_0), C(1, y_0)$ we find
\be 
\label{cxy}
C(x, y_0) = \frac{C(1, y_0) \Gamma(1 + y_0) \Gamma(x)}{\Gamma(x + y_0)}.
\ee
For $y_0 = 1$, \eqref{cxy} implies
\be
C(x, 1) \Gamma(1 + x) = \Gamma(x) C(1, 1). 
\ee
Now replacing $x$ with $y_0$ and using symmetry in the arguments of $C(x,y)$ we deduce that
\be
C(x, y_0) = C(1, 1) \frac{\Gamma(y_0) \Gamma(x)}{\Gamma(x + y_0)} = C(1, 1) B(x, y_0). 
\ee  
For the purposes of determining the OPE coefficients, $C(1,1)$ is often fixed by the leading soft theorems. The holomorphicity  condition is obeyed by celestial amplitudes whose  momentum space behavior is known from soft theorems to be no more singular than a simple pole in frequency. Boundedness in the strip is expected to be inherited from momentum-space amplitudes with sufficiently good UV behavior. Related properties were pointed out in \cite{Guevara:2019ypd}. The argument can be easily generalized to functions which are not symmetric under $\Delta_1 \leftrightarrow \Delta_2$.

\end{document}